\documentclass[twocolumn]{aastex701}

\usepackage{amsmath}
\usepackage{booktabs} 
\usepackage{tabularx}   
\usepackage{xspace}
\usepackage{comment}
\usepackage{enumitem}

\newcommand{\Msun}{\ensuremath{M_\odot}}
\newcommand{\Rsun}{\ensuremath{R_\odot}}

\newcommand{\Zsun}{\ensuremath{Z_\odot}}
\newcommand{\posydon}{\texttt{POSYDON}\xspace}
\newcommand{\mesa}{\texttt{MESA}\xspace}
\newcommand{\Hecore}{\texttt{He core}\xspace}
\newcommand{\COcore}{\texttt{C/O core}\xspace}

\shorttitle{Double Neutron Star Delay Times Across Metallicities}
\shortauthors{Chattaraj et al.}

\begin{document}

\title{Double Neutron Star Delay Times Across Cosmic Metallicities: \\ The Role of Helium Star Progenitors}

\author[0000-0002-6064-388X]{Abhishek Chattaraj}\email{a.chattaraj@ufl.edu}
\affiliation{Department of Physics, University of Florida, 2001 Museum Rd, Gainesville, FL 32611, USA}

\author[0000-0001-5261-3923]{Jeff J. Andrews}\email{jeffrey.andrews@ufl.edu}
\affiliation{Department of Physics, University of Florida, 2001 Museum Rd, Gainesville, FL 32611, USA}
\affiliation{Institute for Fundamental Theory, 2001 Museum Rd, Gainesville, FL 32611, USA}

\author[0000-0002-6842-3021]{Max\,Briel}\email{max.briel@unige.ch}
\affiliation{Département d’Astronomie, Université de Genève, Chemin Pegasi 51, CH-1290 Versoix, Switzerland}
\affiliation{Gravitational Wave Science Center (GWSC), Université de Genève, CH1211 Geneva, Switzerland}

\author[0000-0003-1474-1523]{Tassos\,Fragos}\email{Anastasios.Fragkos@unige.ch}
\affiliation{Département d’Astronomie, Université de Genève, Chemin Pegasi 51, CH-1290 Versoix, Switzerland}
\affiliation{Gravitational Wave Science Center (GWSC), Université de Genève, CH1211 Geneva, Switzerland}

\author[0000-0001-6692-6410]{Seth\,Gossage}\email{seth.gossage@northwestern.edu}
\affiliation{Center for Interdisciplinary Exploration and Research in Astrophysics (CIERA), Northwestern University, 1800 Sherman Ave, Evanston, IL 60201, USA}
\affiliation{NSF-Simons AI Institute for the Sky (SkAI),172 E. Chestnut St., Chicago, IL 60611, USA}

\author[0000-0001-9236-5469]{Vicky\,Kalogera}\email{vicky@northwestern.edu}
\affiliation{Center for Interdisciplinary Exploration and Research in Astrophysics (CIERA), Northwestern University, 1800 Sherman Ave, Evanston, IL 60201, USA}
\affiliation{NSF-Simons AI Institute for the Sky (SkAI),172 E. Chestnut St., Chicago, IL 60611, USA}
\affiliation{Department of Physics and Astronomy, Northwestern University, 2145 Sheridan Road, Evanston, IL 60208, USA}

\author[0000-0003-1749-6295]{Philipp\,M.\,Srivastava}\email{philipp.msrivastava@northwestern.edu}
\affiliation{Center for Interdisciplinary Exploration and Research in Astrophysics (CIERA), Northwestern University, 1800 Sherman Ave, Evanston, IL 60201, USA}
\affiliation{NSF-Simons AI Institute for the Sky (SkAI),172 E. Chestnut St., Chicago, IL 60611, USA}
\affiliation{Electrical and Computer Engineering, Northwestern University, 2145 Sheridan Road, Evanston, IL 60208, USA}

\author[0000-0003-0420-2067]{Elizabeth\,Teng}\email{elizabethteng@u.northwestern.edu}
\affiliation{Center for Interdisciplinary Exploration and Research in Astrophysics (CIERA), Northwestern University, 1800 Sherman Ave, Evanston, IL 60201, USA}
\affiliation{NSF-Simons AI Institute for the Sky (SkAI),172 E. Chestnut St., Chicago, IL 60611, USA}
\affiliation{Department of Physics and Astronomy, Northwestern University, 2145 Sheridan Road, Evanston, IL 60208, USA}
% \collaboration{all}{The POSYDON Collaboration}

\begin{abstract}

Metallicity can play a significant role in massive binary evolution through its impact on the opacity within stellar interiors and wind-driven mass loss. 
In this work, we investigate how the double neutron star (DNS) delay time distribution (DTD) is shaped by the metallicity-dependent evolution of the helium star$-$NS progenitor system.
Drawing from insights rooted in single and binary star physics, we argue that at a given metallicity, the stellar radius during the helium main-sequence sets a lower limit on the size of the DNS orbit at birth.
We then perform population synthesis with the detailed binary evolution code \posydon to illustrate the resulting DTD across a range of metallicities.
Our results indicate that, independent of the common envelope efficiency and reasonable natal kicks, the majority of DNS mergers across metallicities occur typically no earlier than $\simeq 40\,\rm{Myr}$ after star formation and peaks strongly between $80-250\,\rm{Myr}$.
Roughly $15\%$ of DNSs merge within 80 Myr, which may explain $r$-process enrichment in environments with brief star formation histories, while $\gtrsim 20\%$ merge on delay times $>1~$Gyr, providing an explanation for short gamma-ray bursts in old, metal-poor galaxies.
The shape of the DTD can be complex, with a metallicity-dependent split in the dominant formation channel imprinting a characteristic double-peaked structure.
Although ideally oriented natal kicks can produce very short merging DNS, we find that the required kick magnitudes are inconsistent with observations.
Our work has implications for assessing the contribution of DNS mergers to $r$-process enrichment and gamma-ray bursts/kilonovae transients across cosmic time.

\end{abstract}


\section{Introduction} \label{sec:intro}

Since the landmark detection of GW170817 \citep{LIGOGW170817+2017}, we are witnessing a new era of multimessenger astronomy.
In addition to the gravitational waves detected by LIGO-Virgo-KAGRA, GW170817 was accompanied by electromagnetic emission across gamma-ray, X-ray, ultraviolet, optical, infrared and radio wavelengths \citep[see][for a review]{Metzger2017}. These observations revealed a kilonova \citep{LIGOGW170817+kilonova} powered by thermal radiation from the radioactive decay of heavy $r$-process elements synthesized in the merger ejecta \citep{Metzger+2010, Kasen+2017}, providing the first direct observational evidence of $r$-process nucleosynthesis in double neutron star (DNS) mergers. A key ingredient in interpreting DNS mergers within the broader astrophysical and cosmological context is their delay time distribution (DTD), which describes the time between binary formation and merger. The DTD determines whether DNS mergers trace the cosmic star formation history \citep[e.g.,][]{Broekgaarden+2022}, their relative contribution in young and old stellar populations \citep[e.g.,][]{Chu+2022}, and their role in $r$-process enrichment \citep[e.g.,][]{Kobayashi+2023}.

Compact object mergers involving neutron stars (NSs) have been proposed as sites for $r$-process nucleosynthesis since \citet{Lattimer&Schramm1974} and \citet{Symbalisty&Schramm1982}, and remains the best studied source to date\footnote{For other proposed astrophysical sites of $r$-process enrichment, see \citet[][]{Soker2026} for a recent review.}.
Despite being an established site of $r$-process nucleosynthesis, the DNS merger model faces several challenges for explaining all or even most of the $r$-process budget \citep[see][and references therein]{Siegel2019}. Natal kicks imparted to newborn DNSs \citep{Lyne&Lorimer1994} can produce large spatial offsets \citep{Bonetti+2019} or eject them entirely from shallow potential wells characterized by escape velocities of $\mathcal{O}(10\rm\,km/s)$ \citep[see however,][]{Beniamini+2016, Safarzadeh+2019}. Chemical evolution studies have also raised questions about whether DNS mergers can account for the early $r$-process enrichment observed in metal-poor stars in the Galactic halo \citep{Sneden+1994, McWilliam+1995, Hill+2002, Bandyopadhyay+2024}, dwarf galaxies \citep{Shetrone+2001, Ji+2016_Nature, Ji+2016, Hansen+2017, Matsuno+2021, Reggiani+2021, Limberg+2024, Henderson+2025a, Okada+2026}, globular clusters \citep{Roederer+2011, Roederer+2016, Zevin+2019, Kirby+2023, Henderson+2025b}, and the Milky Way disk \citep{Hotokezaka+2018, Kobayashi+2023, Saleem+2025}. The short star formation histories of some of these environments, together with their low metallicities, require that 
whatever the dominant $r$-process site is, it must be active
promptly after star formation, with delay times $\lesssim 10-100 \rm{\,Myr}$.

An alternate way to constrain the delay time distribution is by studying the star-formation histories and redshift distributions of short gamma-ray burst (sGRB) host galaxies \citep{Belczynski+2006, Berger+2007, Zheng&Ramirez-Ruiz2007, Fong+2013, Hao&Yuan2013, Behroozi+2014, Anand+2018, Simonetti+2019, McCarthy+2020,  Zevin+2022, Nugent+2022}. Additionally, the population of DNSs in the Milky Way has been used under the assumption that the observed Galactic sample can be extrapolated to the Universe \citep{Beniamini&Piran2016, Beniamini&Piran2019}. Overall, these studies suggest a wide range of delay times, from several hundred Myr to a few Gyr, although the exact shape of the DTD, which conceals the complex, underlying DNS evolution, remains uncertain.

In this work, we have adopted a theory-driven approach to constraining the DNS DTD across a cosmological range of metallicities based on detailed evolutionary models of DNS formation. 
Previous studies of isolated binary evolution leading to the formation of DNS systems \citep[see][and references therein]{Tauris+2017} have established a standard evolutionary scenario. After the primary star in a massive stellar binary evolves into a neutron star, the principal formation channel \footnote{While this remains the dominant channel at solar metallicity, variations to this channel may be relevant at low metallicities \citep[e.g.,][]{Chruslinska+2018, Stevance+2023}.} involves a common envelope (CE) phase \citep{Ostriker1973, Paczynski+1976} where the now more-massive secondary star unstably overfills its Roche lobe onto the less-massive, first-born NS. At solar metallicity, detailed binary evolution models show that this phase bifurcates depending on the donor's evolutionary stage, producing two distinct subpopulations of DNSs, only one of which merges in a Hubble time \citep{Chattaraj+2026}. The two subpopulations are broadly consistent with the sample of Galactic DNSs \citep{Andrews&Mandel2019}, highlighting the importance of detailed binary evolution modeling in understanding DNS formation. 

Upon exiting the CE, the secondary star now expands into a helium giant, which leads to an additional round of mass transfer (MT), the so-called Case BB MT phase \citep{Delgado&Thomas1981, Ivanova+2003, Dewi&Pols2003}. Eventually the secondary collapses in a supernova (SN), which can impart a combination of a Blauuw kick due to instantaneous mass loss \citep{Blaauw1961} and a natal kick due to small asymmetries in the SN \citep{Lyne&Lorimer1994}. If the binary remains bound, DNSs in sufficiently tight orbits will go on to merge in a Hubble time, emitting gravitational radiation \citep{LIGOGW170817+2017, LIGOGW190425+2020}. 

The He star$–$NS binary thus constitutes the final evolutionary phase prior to DNS formation.
The evolution of the He star, together with ensuing mass transfer and the SN kick, can influence the properties of the DNS system in non-trivial ways  \citep{Dewi+2002, Ivanova+2003, Tauris+2015, Andrews&Zezas2019, Jiang+2024, Nair+2025, Chattaraj+2026}, making this phase critical for understanding the DTD. 
%making its evolution critical for understanding the DTD. 
In this paper, we shall address two key questions: (i) how does the evolution of the He star$–$NS system vary with metallicity and influence the DNS birth orbit; and (ii) what robust features of the DTD can be inferred from a comprehensive modeling of DNS evolution across a cosmological range of metallicities?
We use the detailed binary evolution code \posydon \citep{Fragos+2023, Andrews+2025} to tackle these questions. 

This paper is structured as follows. In Section \ref{sec:physics_expectations}, we identify the dominant factors setting the orbital separation at DNS birth, and therefore, the merger time, focusing on the metallicity-dependent evolution of the He star$–$NS binary. Section \ref{sec:pop_synth} presents a full-scale population synthesis study with the \posydon suite, simulating binaries from two H-rich stars through DNS formation and merger, and discussing the implications of the resulting DTD across a cosmological range of metallicities. 
Finally, Section \ref{sec:conclusions} summarizes the conclusions of this work and highlights relevant caveats.

\section{Physical expectations from metallicity-dependent single and binary star evolution}
\label{sec:physics_expectations}

We use the 1-D stellar evolution code \mesa \citep[][]{Paxton+2011, Paxton+2013, Paxton+2015, Paxton+2018, Paxton+2019, Jermyn+2023} to model the evolution of single He stars and He star$-$NS binaries at eight metallicities, $Z=$ [$10^{-4}$, $10^{-3}$, $10^{-2}$, $0.1$, $0.2$, $0.45$, $1$, and $2$] $Z_\odot$, with $\Zsun=0.0142$, examining how the underlying stellar physics shapes the orbital properties of DNS systems at birth. Our simulation setup uses \mesa r11701 and is identical to that of \cite{Andrews+2025}. 

\subsection{Radial expansion of the helium star}
\label{subsec:he_star_radius}

\begin{figure}
    \centering
    \includegraphics[width=1.00\linewidth]{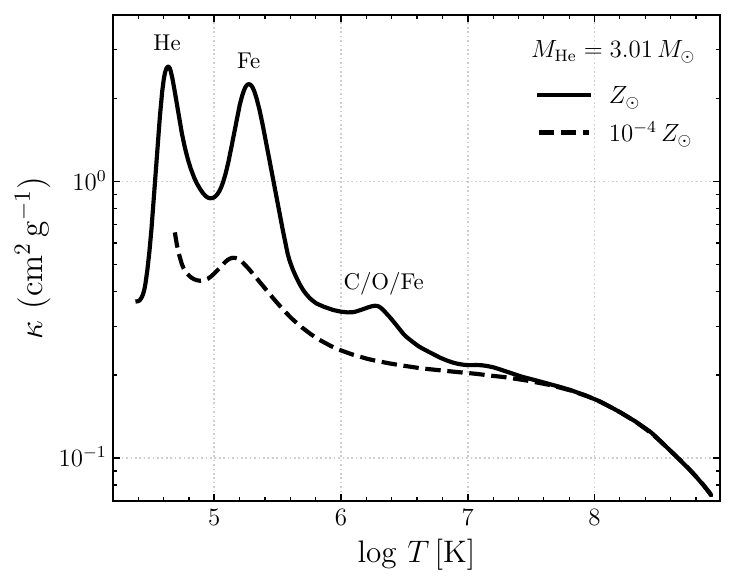}
    \caption{Radiative opacity as a function of temperature in the interior of a $\sim3\Msun$ He star at core carbon exhaustion, shown for two initial metallicities $Z=\Zsun$ (solid) and $Z=10^{-4}\Zsun$ (dashed). The larger opacity bumps for the $Z=\Zsun$ model are indicative of increased radial expansion and higher wind mass loss rates.}
    \label{fig:opacity}
\end{figure}

Although it is not the only effect, we first consider how metallicity influences the radiative opacity within the He star (we consider its affect on winds in Section~\ref{subsec:binary_rlo}).
In Figure~\ref{fig:opacity} we show the opacity as a function of temperature for two He star models with identical masses but different metallicities, $Z=\Zsun$ (solid) and $Z=10^{-4}\Zsun$ (dashed), evaluated at the final evolutionary profile (core-carbon exhaustion). At the high temperatures of the stellar core, the opacity is dominated by Thomson scattering \citep{Kippenhahn&Weigert1994} which is largely insensitive to metallicity. In contrast, the envelope opacity depends strongly on the composition, with bound-free and bound-bound transitions producing characteristic opacity bumps at $\log\,T \simeq 4.7$ (He), $\log\,T \simeq 5.3$ (Fe), and $\log\,T \simeq 6.3$ (C/O/Fe) \citep{Cox&Tabor1976, Iglesias&Rogers1991, Iglesias+1992, Cantiello+2009}. These opacity features are significantly more pronounced at $Z=\Zsun$, with the C/O/Fe bump absent in the $Z=10^{-4}\Zsun$ model.

\begin{figure*}
    \centering
    \includegraphics[width=0.95\linewidth]{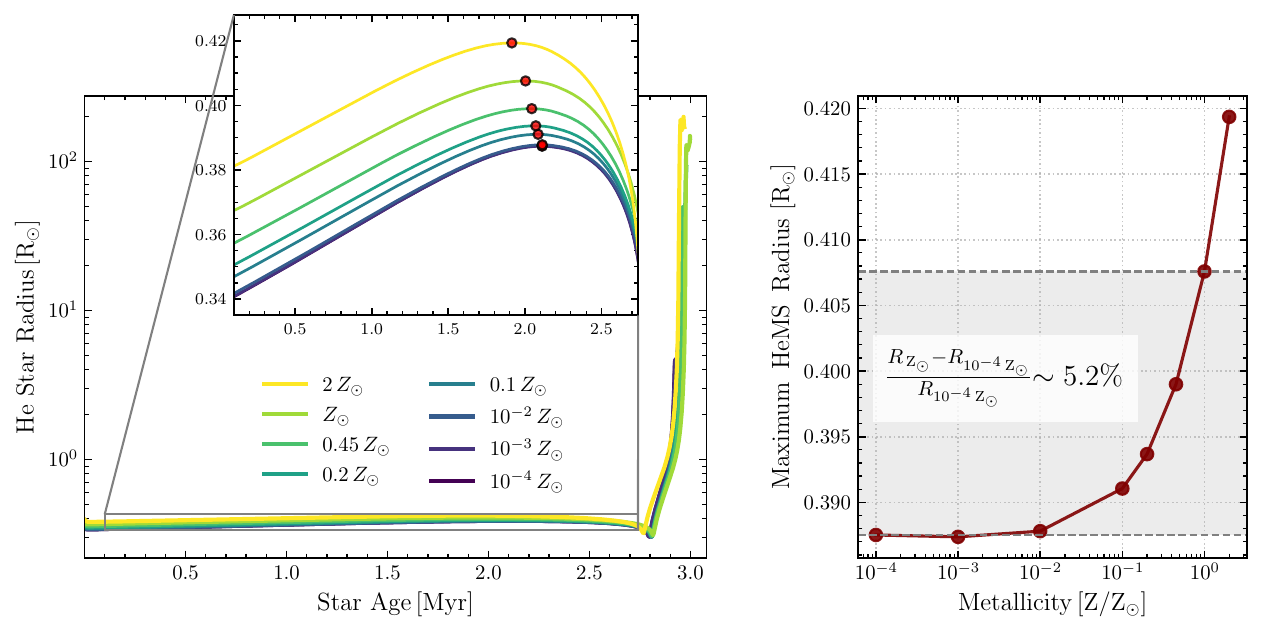}
    \caption{Radial expansion of a 2.3\,\Msun\,He star (representative mass) at different metallicities with a zoomed-in inset on the He main sequence phase. Although the radius remains nearly constant (note the y-axis range in the inset), its exact value varies with metallicity. We pick the maximum radius attained during this phase for each metallicity (red dots) and show the resulting trend on the right panel. The $\Zsun$ model remains roughly 5\% larger compared to the $10^{-4}\Zsun$ model through the HeMS phase.}
    \label{fig:He_star_radius}
\end{figure*}

The enhanced opacity at higher metallicity indicates envelope inflation \citep[e.g.,][]{Sanyal+2017, Xin+2022}, introducing a metallicity dependence in the radial evolution of He stars. Although the effect is modest, lower metallicity stars, with their reduced opacities, remain more compact throughout their long-lived He main-sequence (HeMS) phase. The left panel of Figure~\ref{fig:He_star_radius} illustrates this trend, showing that He stars evolve with systematically smaller radii during the main sequence at lower metallicities. The maximum HeMS radius reached at each metallicity is marked by red dots and shown in the right panel, revealing a clear decrease with decreasing metallicity. 
 
This metallicity-dependent radial expansion levies a  lower limit on the size of the NS$-$He star orbit. Physically, the orbital separation must exceed the sum of their radii, i.e. $a > R_{\rm He} + R_{\rm NS} \simeq R_{\rm He}$; although a more realistic limit can be derived from the Roche lobe radius (see Section \ref{subsec:binary_rlo}). Thus, the smallest possible pre-SN separation is a factor of order unity times the size of the He star. As we show in Sections~\ref{subsec:binary_rlo} and~\ref{subsec:sn_kicks}, although mass transfer and the SN explosion modify the orbit, the pre-SN separation, and therefore the He star radius, remains a reliable predictor of the orbital separation at DNS birth. 
Consequently, progenitor systems that are otherwise identical but have different He star metallicities will form DNSs with systematically different orbital separations, set by the metallicity-modulated HeMS radius.

Assuming the DNS orbital separation at birth is $a_{\rm i}$, the inspiral time $t_{\rm m}$ for a circular orbit scales as $t_{\rm m} \propto a_{\rm i}^4$, making it sensitive to even small changes in the orbital separation. Considering two systems with initial separations $a_1$ (low metallicity) and $a_2 = a_1 + \delta a$ (higher metallicity), and $\delta a \ll a_1$, expanding $t_{\rm m}(a)$ about $a_1$ to first order gives

\begin{align}
t_{\rm m}(a_1 + \delta a) 
&\approx t_{\rm m}(a_1) \left(1 + 4 \frac{\delta a}{a_1} \right) \nonumber \\
\Rightarrow \quad
\frac{\delta t_{\rm m}}{t_{\rm m}} 
&\approx 4 \frac{\delta a}{a_1}
\end{align}

From the right panel of Figure \ref{fig:He_star_radius}, the maximum HeMS radius at $\Zsun$ is roughly 5\% larger than at $10^{-4}\Zsun$, suggesting an increase in inspiral time of $\simeq 20\%$. Thus, even modest metallicity-driven differences in the He star radii can translate into potentially observable variations in DNS merger times.

\subsection{Orbital evolution before and during RLO}
\label{subsec:binary_rlo}

\begin{figure*}
    \centering
    \includegraphics[width=0.95\linewidth]{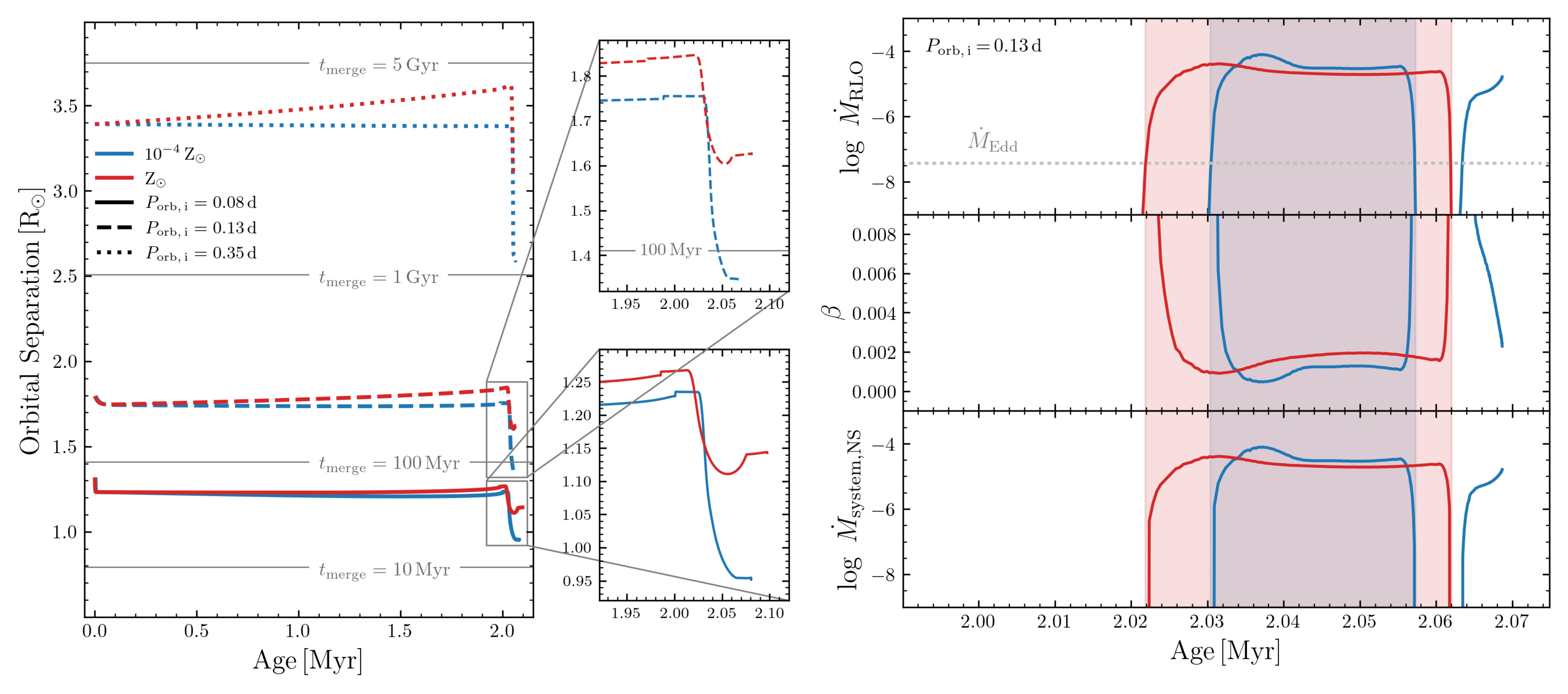}
    \caption{Evolution of a $2.9\Msun$ He star and a $1.43\Msun$ NS. (\textit{Left panel}) Evolution of the binary separation for three different initial orbital periods (different linestyles) and at two metallicities: $\Zsun$ (red) and $10^{-4}\Zsun$ (blue). Horizontal gray lines show the inspiral times for a circular binary with two $1.4\Msun$ NS components born at the corresponding separations. (\textit{Right panel}) Mass transfer parameters for the binary with $P_{\rm orb,i}=0.13\rm d$, showing (from top to bottom) the mass transfer rate, the accretion efficiency, and the systemic mass loss rate via isotropic re-emission from the accretor. Shaded regions mark RLO. For otherwise identical binary configurations, the lower metallicity system experiences weaker wind mass loss-driven orbital widening prior to RLO, and shrinks its orbit more efficiently during RLO, leading to more compact pre- and post-RLO separations. 
    }
    \label{fig:orbital_evolution}
\end{figure*}

Figure~\ref{fig:orbital_evolution} shows the orbital evolution of a $2.9\Msun$ He star and a $1.43\Msun$ NS binary computed with \mesa. The left panel illustrates the evolution of the binary separation for three otherwise identical systems, differing only in their initial orbital periods (indicated by different linestyles) and metallicity: $\Zsun$ (red) and $10^{-4}\Zsun$ (blue). In addition to modulating its opacity, metallicity in the stellar atmosphere affects the wind mass loss rate of the He star \citep{Castor+1975}, influencing the outcome during MT through its impact on the orbit. We model He star winds following the Wolf–Rayet wind prescription from \citet{Nugis&Lamers2000}, with the mass loss rate scaling with surface metallicity as $\dot{M}\propto\sqrt{Z_{\rm surf}}$. As a result, higher metallicity He stars experience stronger wind-driven mass and angular momentum loss, leading to greater orbital widening during core He burning, prior to RLO.

Upon core He exhaustion, the He star expands its envelope and fills its Roche lobe, initiating mass transfer onto the NS. Across identical He star$-$NS systems, we find that binaries at a lower metallicity (with already tighter pre-RLO orbits) experience more efficient orbital shrinkage during RLO, leading to even shorter post-RLO separations. On the right panel of Figure~\ref{fig:orbital_evolution}, we show the detailed mass transfer evolution for the system with initial period $P_{\rm orb,i}=0.13\rm\,d$, and shaded regions indicating RLO. The He star dumps mass rapidly onto the NS (see the highly super-Eddington mass transfer rates in the top row), which accretes only a tiny fraction of the transferred material (second row). The rest is expelled from the vicinity of the NS via isotropic re-emission (third row). This mass is assumed to leave the binary at the specific angular momentum of the NS, causing the orbit to shrink significantly. At higher metallicities, stronger wind mass loss partially counteracts this shrinkage by simultaneously widening the orbit, whereas this effect is largely negligible at lower metallicities. This effect remains sensitive to uncertainties in modeling He star wind mass loss rates (see Appendix \ref{subsec:he_star_wind}). The horizontal gray lines in the left panel, indicating inspiral times for a circular binary with two $1.4\Msun$ NS components formed at the corresponding separation, illustrate that for otherwise identical initial binary configurations, lower metallicity leads to more compact orbits and shorter inspiral times. 

In Section \ref{subsec:he_star_radius}, we argued that the minimum pre-RLO orbital separation is set by the size of the He star. During RLO, the He star remains in close Roche contact ($R_{\star} \simeq R_{\rm L}$), and since $R_{\rm L} = a f(q)$ \citep{Eggleton1983}, the orbital separation continues to track the He star radius. Thus, even after mass transfer concludes, the He star radius remains a reliable predictor of the binary separation. In the next section, we demonstrate that this conclusion holds largely true even after the supernova explosion, provided the binary remains bound.

\subsection{Dynamics of the supernova kick}
\label{subsec:sn_kicks}

\begin{figure*}
    \centering
    \includegraphics[width=1.00\linewidth]{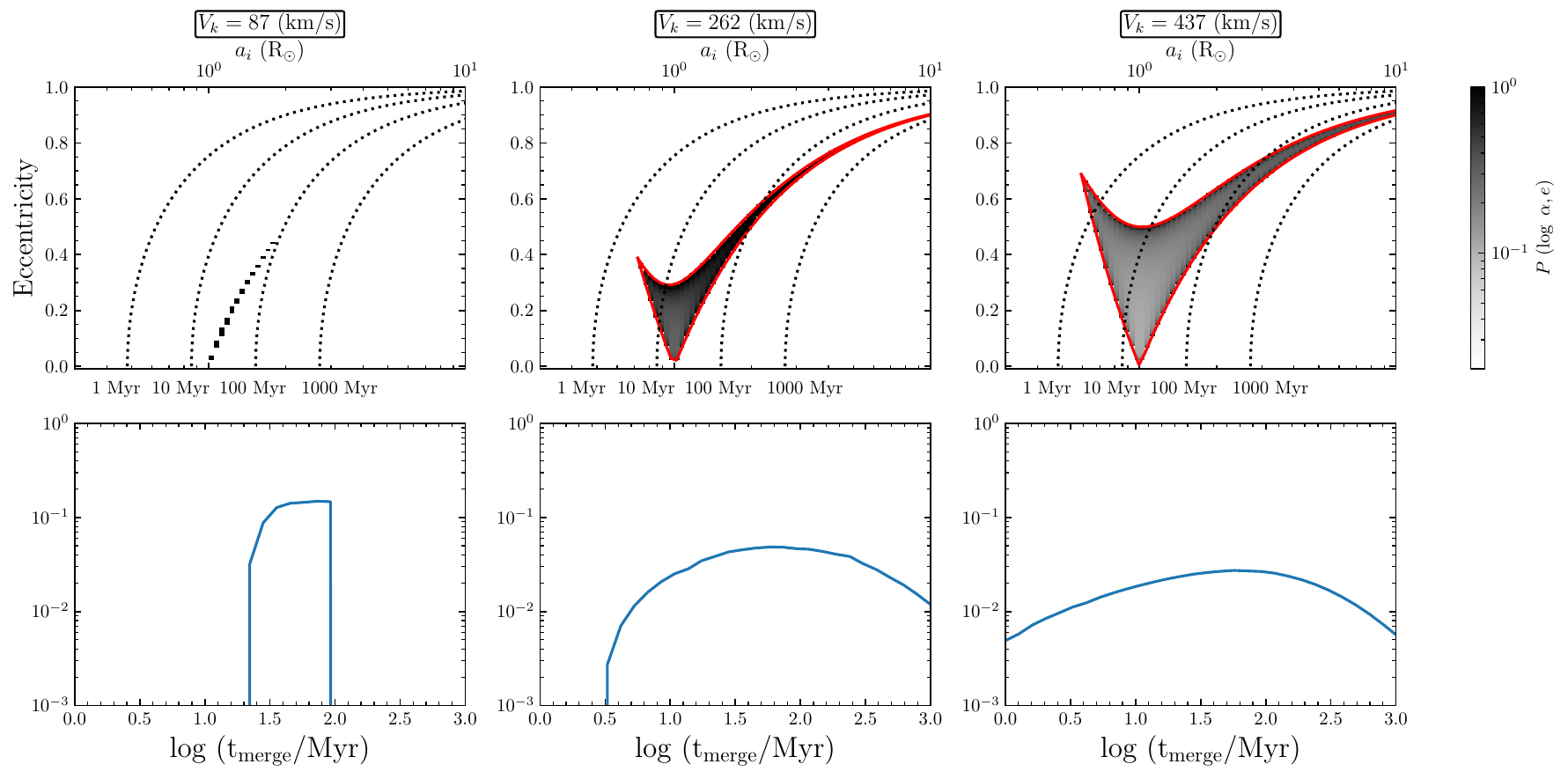}
    \caption{(\textit{Top row}) Distribution of post-SN orbital separations and eccentricities for a fixed pre-collapse He star mass $M = 1.8\Msun$ and a range of natal kick magnitudes, $V_{\rm k}=[0.1,\,0.3,\,0.5]\,V_{\rm orb}$ (left to right). Black dotted lines indicate constant merger times. The distributions are confined within the analytic limits shown by the red lines (see discussion in Section \ref{subsec:sn_kicks}), with high density regions near the boundaries. While in the majority of cases the natal kick widens the binary; some systems are kicked into a smaller orbit. (\textit{Bottom row}) Although larger natal kicks tend to broaden the distribution of merger times, the leftmost panel, with a natal kick $<$100 km/s, is most representative of the observed DNS population in the Milky Way. 
    }
    \label{fig:post_SN_orbit}
\end{figure*}

As the He star exhausts its nuclear fuel, it collapses in a supernova explosion, imparting a kick to the binary \citep{Blaauw1961, Lyne&Lorimer1994}. The kick modifies the orbit and sets its eccentricity and orbital period, thus determining the DNS merger time. To quantify the dynamical impact of the SN explosion, we compute the post-SN orbital distributions assuming an isotropically oriented kick of fixed magnitude. The post-SN orbit is characterized by the eccentricity $e$ and the non-dimensional quantity $\alpha = a_{\rm f}/a_{\rm i}$, defined as the ratio between the post- and pre-SN orbital separation. For more details, we refer the reader to \citet{Kaolgera1996} and \citet{Andrews&Zezas2019}. 

The top row of Figure \ref{fig:post_SN_orbit} shows the resulting distributions for a fixed pre-SN separation $a_{\rm i} = 1\Rsun$ and He star mass $M_{\rm He,\,i} = 1.8\Msun$, representing the mass immediately prior to core collapse. From left to right, the panels correspond to natal kick magnitudes of 10\%, 30\%, and 50\% of the pre-SN orbital velocity. The post-SN binaries populate a well-defined region of the parameter space, bounded by the red curves showing the analytic limits, $(1+e)^{-1}<\alpha<(1-e)^{-1}$ \citep[first identified by][]{Flannery&vandenHeuvel1975}, and the maximum allowed post-SN eccentricity \citep[see Eq.~16 in][]{Andrews&Zezas2019}. The distribution is densely populated near these boundaries, indicating that even isotropic kicks do not produce arbitrary post-SN orbits. Increasing the kick magnitude shifts the distribution towards wider and more eccentric systems (right branches), but it also increases the fraction of binaries that experience orbital shrinkage (left branches in the second and third panels). The latter occurs when a kick comparable to the Keplerian velocity of the orbit (typically $\sim \mathcal{O}(100)~\mathrm{km/s}$ at $a_{\rm i}=1\Rsun$) is directed opposite to the orbital motion, sharply reducing the angular momentum. If the binding energy of the post-SN system remains larger than its kinetic energy, this can produce a DNS in a very tight orbit \citep[see][]{Beniamini&Piran2024}. Using the equations in \citet{Peters1964}, we compute the corresponding merger times, shown in the bottom row. As expected, SN kicks broaden and partially randomize the merger time distribution.

Crucially, although some systems shrink due to the SN, the dominant outcome is a post-SN orbit that is comparable to or wider than the pre-SN orbit. Moreover, while the random orientations guarantee that some systems receive kicks directed opposite to the orbital motion, previous studies find that typical magnitudes are $\lesssim50 \rm{km/s}$ \citep{Guo+2024, Disberg+2024, Chattaraj+2026}, which is insufficient to halt any binary motion. Therefore, the allowed post-SN orbits remain largely confined to a region whose lower bound is set by the pre-SN separation. Since the pre-SN separation is set by the He star’s radius during the HeMS phase (Sections \ref{subsec:he_star_radius} and \ref{subsec:binary_rlo}), this establishes a robust link between the He star radius and the post-SN orbital separation, and therefore the resulting DNS merger time. Thus, despite the stochastic nature of SN kicks, the size of the He star remains the primary predictor of the DNS merger time.

\subsection{A lower limit on the NS$-$NS inspiral time}
\label{subsec:lower_limit}

As discussed in Section \ref{subsec:binary_rlo}, mass transfer cannot shrink the orbit below the donor's Roche lobe, which acts as a floor to the binary separation. Any further tightening would cause the donor to substantially overfill its Roche lobe, triggering a dramatic increase in the mass transfer rate through L1, which is extremely sensitive to the relative radius excess, $\delta R_{\rm d} = (R_{\rm d}-R_{\rm L})/R_{\rm L}$ \citep[e.g.,][]{Cehula&Pejcha2023}. We therefore take $R_{\star} \simeq R_{\rm L}$ as the minimum birth separation of the NS$–$NS pair when estimating the minimum inspiral times.

We highlight two underlying assumptions. First, we preclude unstable mass transfer between the He star and the NS as a possible way to produce DNSs. Our solar metallicity models have previously shown that such evolution typically leads to either a CE merger or a stripped remnant too low in mass to form a NS \citep{Chattaraj+2026}; we find this result to hold across all metallicities explored here (\citealt{Chattaraj+2026_in-prep}).
Second, we assume that the post-SN orbit remains comparable to or wider than the pre-SN orbit, a reasonable approximation as discussed in Section \ref{subsec:sn_kicks}. Under these assumptions, we estimate minimum NS$-$NS inspiral times of $t_{\rm insp, \, min} \simeq\mathcal{O} \rm{(10^5~yr)}$ assuming a circular orbit.
Since the inspiral time scales with eccentricity as $(1-e^2)^{7/2}$, $t_{\rm insp, \, min}$ decreases by roughly a factor of 3 for $e \approx 0.5$, and can be as short as $\mathcal{O} \rm{(10^3~yr)}$ for $e \approx 0.9$. At solar metallicity, $t_{\rm insp, \, min}$ remains systematically $\sim 20\%$ longer. At such short inspiral times, the delay time is dominated by DNS formation, which is typically $\approx \mathcal{O} \rm{(10~Myr)}$ (see Section \ref{subsec:formation_times}).

\section{Binary Population Synthesis}
\label{sec:pop_synth}

Having examined the impact of the metallicity-dependent evolution of the progenitor binary, we complement our goal of characterizing the DTD by performing full-scale population synthesis, modeling the complete evolution from ZAMS to DNS merger across metallicities using the binary simulations suite \posydon \footnote{The version of the code is identified by the commit hash \texttt{888d89cc5} available at \href{https://github.com/POSYDON-code/POSYDON/}{https://github.com/POSYDON-code/POSYDON/}. The \posydon v2 datasets are available on Zenodo under an open-source Creative Commons Attribution license here: \dataset[doi: 10.5281/zenodo.15194708]{https://zenodo.org/records/15194708}.} \citep{Fragos+2023, Andrews+2025}.

\subsection{Setup}
\label{subsec:pop_methods}

We generate synthetic populations of massive binaries at the eight metallicities described in Section \ref{sec:physics_expectations}. \posydon relies on grids of detailed single and binary star models computed with the 1D stellar evolution code \mesa \citep{Paxton+2011, Paxton+2013, Paxton+2015, Paxton+2018, Paxton+2019, Jermyn+2023}. The binary grids are used to model potentially interacting evolutionary phases and include the \texttt{HMS-HMS} grid (two H-rich stars), the \texttt{CO-HMS} grid (a compact object and a H-rich star), and the \texttt{CO-HeMS} grid (a compact object and a He-rich star, meant to represent a donor stripped of its H envelope). To better resolve the parameter space relevant to DNS formation, we supplement the \texttt{CO-HMS} grid with a densely sampled set of models across all eight metallicities \citep[see Section 2.3 in][for models at solar metallicity]{Chattaraj+2026}. Further details will be presented in \citealt{Chattaraj+2026_in-prep}

We consider two natal kick prescriptions: (i) an asymmetric ejecta (\textit{AsymEj}) kick model that calculates the magnitude from the SN ejecta mass \citep[derived from][]{Janka2017}, with the exact prescription given in Eq.~2 of \citet[][]{Chattaraj+2026}; and (ii) a lognormal distribution with $\mu = 5.60$ and $\sigma=0.68$ following \citet{Disberg&Mandel2025}, a revision of the commonly used Maxwellian distribution of \citet{Hobbs+2005}, which omitted a Jacobian term in its original analysis.
We adopt a burst star formation history, in which all binaries are formed at the same time. For the CE phase, we adopt the \texttt{Two\_Phases\_StableMT} option in \posydon, and vary only the ejection efficiency $\alpha_{\rm CE}$ while keeping the core–envelope boundary fixed at a $30\%$ H-mass fraction. The SN outcomes are determined following \citet{Sukhbold+2016} for core-collapse, and \citet{Tauris+2015} for electron-capture supernova. Additional setups specific to DNS formation are identical to \citet{Chattaraj+2026}. 

We make one significant alteration to the \posydon setup when generating DNS populations, motivated by recent observations of stripped stars by \citet{Gotberg+2023}. These observations suggest that stripped He-rich stars ought to have wind mass-loss rates $\lesssim10^{-9}\,M_\odot\,\rm{yr^{-1}}$. However, \texttt{POSYDON} calculates wind mass-loss rates by extrapolating the WR star prescription from \citet{Nugis&Lamers2000} to the regime of low-mass He stars. When compared to the observations, this produces an overestimate of the winds which, as shown in Figure~\ref{fig:orbital_evolution} and described in the surrounding text, leads to increased orbital expansion. To account for this effect,
we simulated all population models such that the He star$-$NS phase (via the \texttt{CO-HeMS} grid) at all metallicities was artificially set to follow the evolutionary track of an equivalent binary in the $10^{-4}\,Z_\odot$ grid (see Appendix \ref{subsec:he_star_wind} for details). This has the practical effect of limiting orbital expansion due to mass loss for all binaries going through the \texttt{CO-HeMS} grid.
We summarize the population models below. \\
(i) \texttt{MODEL01}: $\alpha_{\rm CE}=1.0$, \textit{AsymEj} kick model.\\ 
(ii) \texttt{MODEL02}: $\alpha_{\rm CE}=2.0$, \textit{AsymEj} kick model.\\
(iii) \texttt{MODEL03}: $\alpha_{\rm CE}=0.5$, \textit{AsymEj} kick model.\\
(iv) \texttt{MODEL04}: $\alpha_{\rm CE}=1.0$, lognormal kick model. \\
For each DNS system, the delay time is defined as $t_{\rm delay} = t_{\rm form} + t_{\rm insp}$, where $t_{\rm form}$ is the formation time from ZAMS to DNS birth, and $t_{\rm insp}$ is the inspiral time.

\begin{figure}
    \centering
    \includegraphics[width=1.0\linewidth]{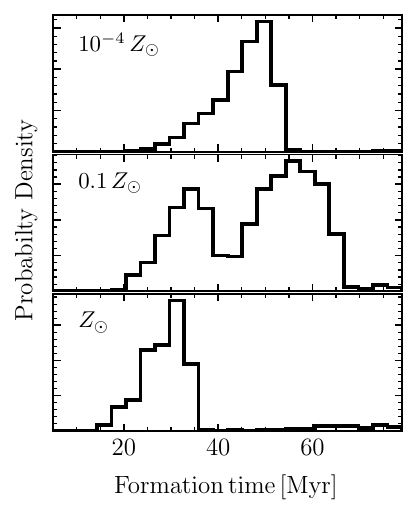}
    \caption{The distribution of formation times (from ZAMS to DNS birth) for merging systems in \texttt{MODEL01} at three metallicities. Formation times scale with the nuclear lifetimes of the H-burning primary stars, which are more massive at higher metallicities, explaining the shorter timescales. The double peak at intermediate metallicities results from the bifurcation of the standard CE formation channel into two subchannels \citep[\Hecore and \COcore;][]{Chattaraj+2026}. A floor at $t_{\rm form} \simeq 20\,\rm{Myr}$ remains consistent across model variations, representing the minimum time after star formation for a DNS system to possibly exist.}
    \label{fig:formation_times}
\end{figure}

\subsection{Distribution of formation times}
\label{subsec:formation_times}

Figure \ref{fig:formation_times} shows the distribution of formation times for merging DNSs in \texttt{MODEL01} at three metallicities, with little variation across population models. The formation times are dominated by the nuclear timescales of the primary H-rich star; and we find that the ZAMS stars at higher metallicities are initially more massive, leading to shorter formation times. Due to the steep mass dependence ($\tau_{\rm nuc} \propto M^{-2.5}$), DNSs form on average $10-15\,\rm{Myr}$ earlier at $\Zsun$ compared to $10^{-4}\,\Zsun$.

At metallicities $\gtrsim 10^{-2}\,\Zsun$, DNS formation proceeds via two subchannels (\Hecore and \COcore) within the dominant CE channel \citep{Chattaraj+2026}. The two subchannels originate from ZAMS progenitors of different masses \citep[see Fig. 17 in][for models at \Zsun]{Chattaraj+2026}, producing a double peak in the $t_{\rm form}$ distribution. However, at $\Zsun$, only the \Hecore channel contributes to merging DNSs, while at $\lesssim 10^{-2}\,\Zsun$, all DNSs form via the \COcore channel, explaining the absence of a second peak in these cases. Crucially, the minimum formation time remains $\simeq 20\,\rm{Myr}$ across all metallicities and model variations, setting the earliest timescale after star formation at which DNSs can possibly appear.

\subsection{Distribution of delay times}
\label{subsec:pop_synth_results}

\begin{figure}
    \centering
    \includegraphics[width=1.0\linewidth]{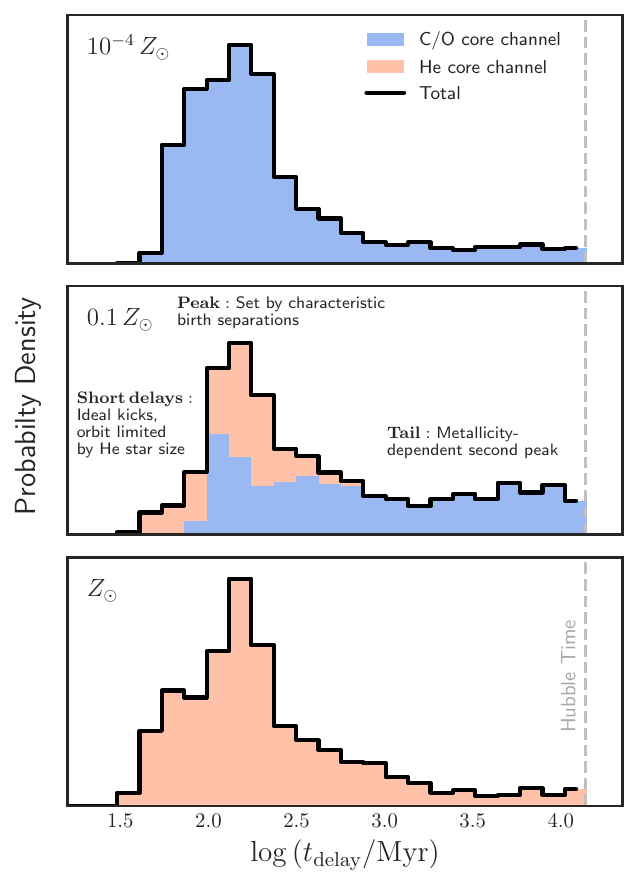}
    \caption{Same as Figure \ref{fig:formation_times} but for delay times. The DTD shape reflects the underlying physics of isolated binary evolution leading to DNS formation. At metallicities $\lesssim 10^{-2}\,\Zsun$, DNS formation proceeds exclusively through the \COcore channel, while only the \Hecore channel produces merging DNS at $\gtrsim \Zsun$. For intermediate metallicites, both channels contribute, producing a characteristic double peak. The accompanying text in the middle panel links the physical processes to the distinct features that give the DTD its shape.}
    \label{fig:dtd}
\end{figure}

\begin{figure*}
    \centering
    \includegraphics[width=0.96\linewidth]{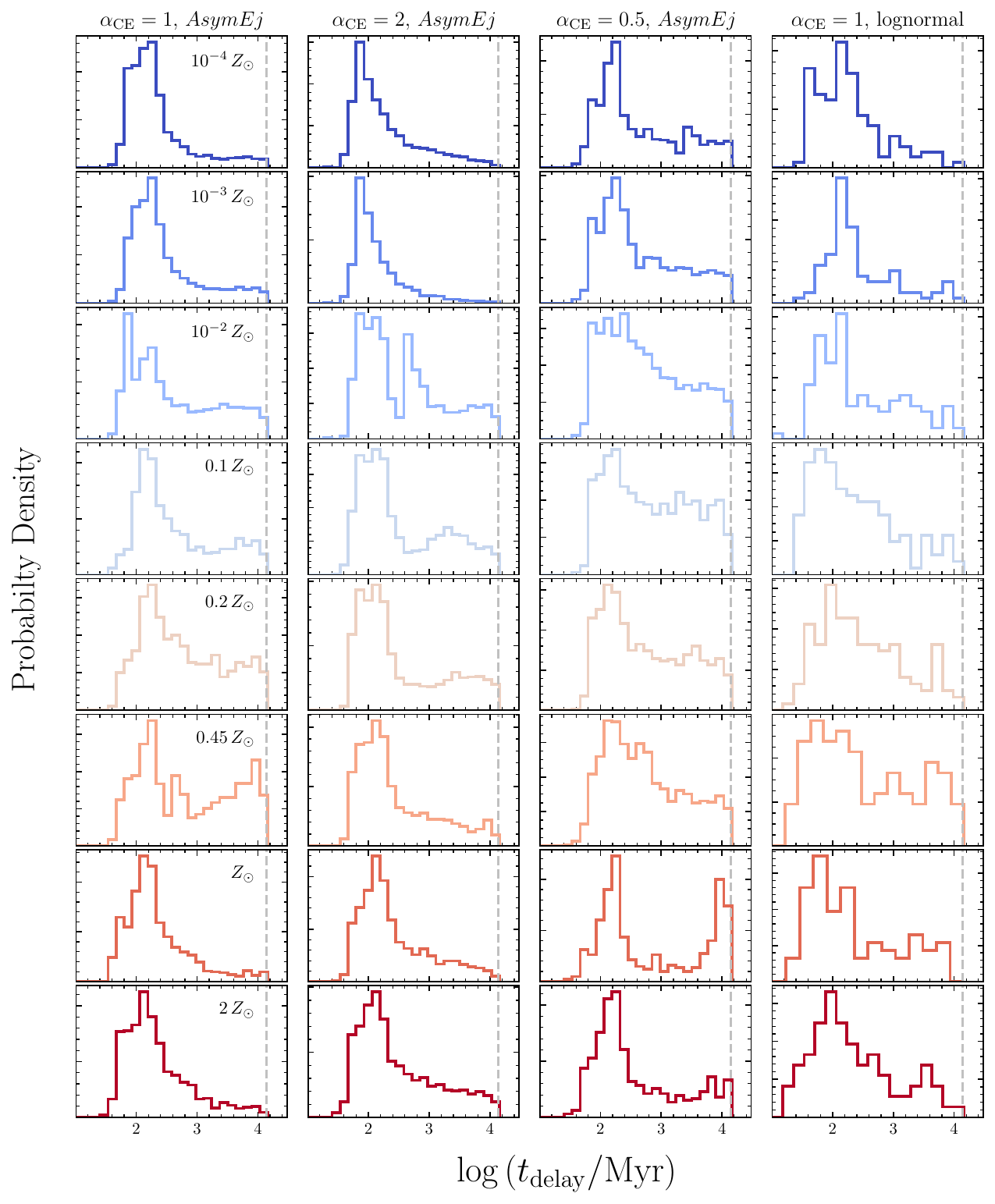}
    \caption{Distributions of DNS delay times from population synthesis across eight metallicities (rows) and four model variations (\texttt{MODELS01-04}; columns), exploring different CE ejection efficiencies ($\alpha_{\rm CE}$) and natal kick prescriptions, while maintaining a fixed core-envelope boundary at $X=0.30$. The gray dotted line marks $t_{\rm Hubble}=13.8\,\rm Gyr$. 
    A bifurcation in the dominant CE formation channel \citep{Chattaraj+2026} produces a secondary peak at some metallicities (see Figure \ref{fig:dtd}).
    Despite model variations that change the overall shape of the DTD, two features remain robust: the majority of DNS mergers occur $\gtrsim 40\,\rm{Myr}$ after star formation and the distribution peaks at $\rm{80-250\,Myr}$ across metallicities.}
    \label{fig:pop_synth}
\end{figure*}

Analogous to Figure \ref{fig:formation_times}, we show the distribution of delay times in Figure \ref{fig:dtd}, highlighting the physical reasons behind the DTD shape. The shortest delay times are dominated by systems with high eccentricities or compact orbits, or both, with a lower limit set by the size of the He star prior to collapse. The DTD peak reflects the characteristic orbital separations at which the majority of merging DNS systems in our populations are formed. The late time behavior is metallicity-dependent: at low ($\lesssim 10^{-2}\Zsun$) and high ($\gtrsim \Zsun$) metallicities, the DTD shows a declining tail, whereas intermediate metallicities show a secondary peak arising due to contribution from a metallicity-dependent bifurcation of the dominant CE channel. Overall, this indicates that the DTD has a complex shape, and a single $t^{-1}$ power law is likely to be an overly simplistic model.

Figure \ref{fig:pop_synth} shows the DTDs for merging DNSs across all eight metallicities (rows) and different model populations (columns; \texttt{MODELS01-04} from left to right).
On average, more systems at low metallicities form with orbits that are compact enough to merge ($\gtrsim 70\%$ of DNSs at $\lesssim 10^{-2}\,\Zsun$ across all models merge in a Hubble time).
While the overall shape of the DTD can vary, two features remain robust and largely independent of the adopted CE treatment and natal kick prescription$-$two assumptions that strongly influence DNS formation \citep{Vigna-Gomez+2018, Chattaraj+2026}.

First, all model populations exhibit a robust minimum to the delay times. Aside from rare outliers (which may have $t_{\rm delay}$ $\simeq 20\,\mathrm{Myr}$), the bulk of DNS mergers begin $\gtrsim 40\,\mathrm{Myr}$ after star formation. This represents the minimum timescales on which isolated binary evolution can produce merging DNSs. Second, the DTD across metallicities consistently peaks between $\simeq 80-250\,\mathrm{Myr}$ depending on model parameters. This trend is largely independent of the explored variations in the binary physics treatment. 
As we move to higher metallicities, particularly at $Z \gtrsim 10^{-2}\Zsun$, the distribution shows a second peak due to the bifurcation into two formation subchannels (\Hecore and \COcore), which disappears at $\simeq \Zsun$ and beyond, for the reasons discussed above (see Figure \ref{fig:dtd}).

Natal kick models that randomly sample from a distribution (e.g., lognormal, rightmost column) dominate the tail at short delay times. The strong kick produces high eccentricities, or when directed opposite to the motion of the He star, effectively shrinks the orbit. Both scenarios shorten the merger time, populating the short delay time end of the distribution. However, as shown in \citet{Chattaraj+2026}, such kick models struggle to reproduce both the wide range of orbital periods observed in the Galactic DNS population, as well as the local merger rate. In contrast, a more physically motivated prescription such as \textit{AsymEj}, where the kick scales with SN ejecta mass, rarely produces such extreme systems. 

Furthermore, we find that applying the $10^{-4}\,\Zsun$ He star wind prescription to all metallicities leads to a reduction in orbital expansion (as described in Section~\ref{subsec:binary_rlo}) of the progenitor binary at high metallicities (see Appendix \ref{subsec:he_star_wind}), increasing the fraction of merging DNSs. However, we find this to have only a perturbative effect on the overall DTD; DNSs systematically maintain wider orbits at high metallicity throughout prior evolutionary phases \citep[e.g.,][]{Gallegos-Garcia+2023}, and our choice of the He star wind prescription has only a minor impact.

\subsection{Discussion}
\label{subsec:discussion}

It is interesting to compare our results with existing literature. The DTD of DNS mergers have been constrained in the past using observations which primarily fall into two camps: $r$-process enriched stars and short GRBs. Chemical evolution studies of the Milky Way suggest that both a prompt and a delayed enrichment channel are required \citep{Hotokezaka+2018, Cote+2019}. \citet{Skuladottir&Salvadori2020} find that a rapid channel with $t_{\rm delay} \lesssim 100\,\rm Myr$ together with a delayed source at $t_{\rm delay} \gtrsim 4\,\rm Gyr$ can reproduce the observed abundance patterns in the Milky Way and its dwarf satellites, while \citet{Chen+2025} find that the prompt channel should roughly trace star formation. 
It has been noted that even if DNS mergers follow a prompt DTD, the subsequent cooling, mixing and incorporation of the ejected material into the next generation of stars could take $\sim500-1000$ Myr \citep{Naidu+2022}, especially if natal kicks eject DNSs out of their galactic hosts. Similarly, \citet[][]{Nugent+2025} find typical enrichment timescales of $t_{\rm min} = 134^{+171}_{-83}$ Myr from observations of GRBs and kilonovae.
Meanwhile, \citet{Kobayashi+2023} find that most population synthesis codes struggle to reproduce observed abundances unless low metallicity DNSs merge rapidly and with substantial ejecta mass.
In addition, \citet{Holmbeck&Andrews2024} report that predicted $r$-process yields from Galactic DNSs underproduce heavy elements relative to the Solar system, though past mergers of low metallicity DNSs could help reconcile this discrepancy.

While our models predict a minimum delay time of $t_{\rm min} \simeq 40\,\rm Myr$ \citep[which is broadly consistent within model-dependent variations in previous population synthesis studies;][]{Dominik+2012, Belczynski+2018_DNSmerger, Chruslinska+2018}, we note that the commonly adopted $t^{-1}$ power-law DTD originates from assuming the same log-flat distribution $dN/da \propto a^{-1}$ for DNS birth separations as O/B stars at ZAMS. Combined with $t \propto a^4$, we have $da/dt \propto t^{-3/4}$. It follows that $dN/da \propto t^{-1/4}$ and $dN/dt \propto t^{-1}$.
However, uncertainties in the evolutionary processes responsible for shrinking the orbit makes this assumption of a log-flat post-SN separation distribution difficult to justify in realistic scenarios of DNS formation. More generally, if the post-SN separations follow a distribution $dN/da \propto a^{-\beta}$, then we obtain a DTD given by $dN/dt \propto t^{-(\beta+3)/4}$; somewhat steeper distributions with $\beta \simeq 3$, lead to $dN/dt \propto t^{-1.5}$ \citep{Belczynski+2018}. While our results suggest that modeling the DTD with a single power law may be overly simplistic, a preliminary fit to the tail following the peak in our results are in agreement with a growing consensus from multiple methods that, when represented as a power law, the DTD must be steeper than $t^{-1}$ \citep[e.g.,][]{Zevin+2022}. 

To date, there are two known ultra-faint dwarf galaxies Reticulum II and Tucana III that show clear signatures of $r$-process enrichment. In Reticulum II, the high $r$-process abundances, inferred event rate \citep{Ji+2016_Nature}, and homogeneous dilution pattern \citep{Ji+2023} are consistent with enrichment from a prompt DNS merger. The similar abundance patterns observed in Tucana III further suggest a common origin for $r$-process-enhanced stars in the Milky Way halo and dwarf galaxies \citep{Hansen+2017}. 

Additionally, independent constraints on the DTD come from sGRBs. Observations of sGRBs in dwarf galaxies provide support for low metallicity DNS mergers \citep{Nugent+2024}. Using the observed sample of sGRBs, \citet{Zevin+2022} inferred a minimum delay time $t_{\rm min} = 184^{+67}_{-79}$ Myr at 99\% credibility \citep[however, see][]{Pracchia&Salafia2026, DeSantis+2026}. Although \citet{Nugent+2022} found that roughly 84\% of sGRB host galaxies are star-forming, accounting for host redshift and stellar age distributions reveal a broad DTD, with a fast-merging channel at $z \gtrsim 1$ and declining DNS formation efficiency toward lower redshifts. 
Alternatively, \cite{Maoz&Nakar2024} derive a DTD from Galactic DNSs, finding that fast-merging systems are well fit with a power-law assuming an exponential cutoff at $\approx 300\,\rm{Myr}$, which comfortably includes our predictions for the peak delay times (Figure \ref{fig:pop_synth}).

Our models indicate that the bulk of DNS mergers occur $\simeq 80-250\,\rm{Myr}$ after star formation -- significantly shorter than the $\rm{Gyr}$-scale delays often inferred in earlier work \citep[][]{Belczynski+2002, Nakar+2006, Hao&Yuan2013, Wanderman&Piran2015, Vigna-Gomez+2018}. 
In particular, we find that a large fraction ($\gtrsim 70\%$) of DNSs formed at low metallicity ($\lesssim 10^{-2}\,\Zsun$) merge within a Hubble time. 
About $15\%$ will go on to merge within 80 Myr, which may be sufficiently short to enrich environments with brief star formation episodes, such as globular clusters.
At the same time, $\gtrsim 20\%$ merge on timescales $>1$ Gyr, which may help explain short GRBs observed in old, metal-poor galaxies \citep[e.g., NGC 4993, the host galaxy of GW170817, is an elliptical galaxy;][]{Im+2017}. While our models demonstrate that DNS mergers can plausibly contribute to enrichment in low metallicity environments, we caution against interpreting this as definitive evidence. A more accurate picture would include a convolution of the cosmic star formation history with the DTD, which is beyond the scope of this work.

\section{Conclusions}
\label{sec:conclusions}

DNS mergers are associated with multi-wavelength observations and contribute to at least part of the observed $r$-process budget, yet their delay times remain poorly constrained. In this work, we address this problem from the theoretical perspective of isolated binary evolution leading to DNS formation and merger. Using detailed stellar models, we identify the evolutionary factors within the progenitor binary that sets the DNS birth separation, determining its merger timescale ($t_{\rm m} \propto a^4$). We complement this analysis with population synthesis using \posydon to study the resulting delay time distributions across a cosmological range of metallicities and reasonable variations in binary physics. Our main conclusions are summarized below:

\begin{itemize}

\item The formation time, which encompasses all evolutionary processes leading to DNS birth, including binary interactions and supernovae, has a robust minimum of $\simeq 20\,\rm{Myr}$. This represents the earliest timescale after star formation at which DNS systems can appear in an environment, and is set by the nuclear timescale of the H-rich primary star.
    
\item The size of the He star during the core-He burning phase determines the orbital scale of the DNS system at birth (formation of the second NS). Suppressed radial expansion and wind mass loss at low metallicity leads to systematically tighter orbits. This trend persists through the subsequent RLO and supernova phase. 

\item Detailed population synthesis shows that the DTD shape can vary with model assumptions and binary physics uncertainties. The dominant CE formation channel splits into two subchannels, \Hecore and \COcore, with only the \Hecore channel producing merging DNSs at $Z \gtrsim \Zsun$, while DNS formation proceeds exclusively via the \COcore channel at $Z \lesssim 10^{-2}\Zsun$. This metallicity-dependence of the formation pathways imprints a characteristic double-peaked structure in the DTD.

\item Two features remain particularly robust across metallicities and variations in the binary physics treatment: a consistent lower bound of $\simeq 40\,\rm{Myr}$ and a pronounced peak around $80-250\,\rm{Myr}$ (Figure \ref{fig:pop_synth}). Together, these represent the minimum and characteristic timescales on which isolated binary evolution produces merging DNSs.

\end{itemize}

Finally, we discuss a few caveats associated with this work. First, we have not exhaustively tested the full range of binary physics uncertainties. Instead, we focused on factors that have been identified in prior studies \citep[e.g.,][]{Vigna-Gomez+2018, Chattaraj+2026} as most critical to DNS formation. Second, \posydon adopts certain physics assumptions related to stellar microphysics (e.g., radiative opacity) and macrophysics (e.g., stellar winds, convection) while constructing the detailed grids of binary evolution models. Although we expect any effects from these changes to be small, a full binary population synthesis study testing alternative prescriptions is required to demonstrate the exact magnitude of possible variations on the DNS DTD.
Third, we are limited by our accuracy in modeling low-mass He star winds. In this work, informed by observations of stripped stars \citep{Gotberg+2023}, we adopted a prescription where the He star winds at all metallicities represent those at $10^{-4}\,\Zsun$ (see Appendix \ref{subsec:he_star_wind}). While this approach mitigates any unphysical orbital expansion, uncertainties in the mass loss rates could have an influence on the pre-SN orbital separations.
Finally, we note that at low metallicity, the reduced internal opacity can allow a donor star to shrink within its Roche lobe before the entire hydrogen envelope is stripped \citep[see also \citealt{Vigna-Gomez+2022}]{Gotberg+2017}. This can introduce inaccuracies in \texttt{POSYDON}'s treatment of the post-CE core as a pure He star with no surface hydrogen. As a result, the post-CE orbital separation distribution, and, in turn, the shape of the resulting DTD, may be affected.

Despite these limitations, our results provide a detailed link between DNS delay times across metallicities and the binary evolution of their progenitor systems.
These model populations offer a foundation for future studies related to DNS mergers and their observable counterparts. Recent work \citep[e.g.,][]{Fishbach+2026, Kunnumkai+2026} has highlighted tensions in the inferred NS-NS merger rates across different cosmic probes. Convolving the DTDs derived in this work with realistic cosmic star formation histories provides independent predictions for the NS-NS merger rates. With additional assumptions, these predictions can be extended to GRBs and kilonovae across the metallicity and redshift space.
Our models can also facilitate estimates for $r$-process yields from DNS mergers \citep[e.g.,][]{Holmbeck&Andrews2024}. We plan to address these applications in future work.
As observations continue to expand the sample of $r$-process enriched, metal-poor stars and the electromagnetic transients associated with DNS mergers, our low-metallicity models can help bridge the gap between binary evolution models and observed populations.

\begin{acknowledgments}
We thank the anonymous referee for their review of the manuscript. We would like to acknowledge useful discussions with Ylva G\"otberg, Rana Ezzeddine, Avrajit Bandyopadhyay and Alexander Ji. The \posydon{} project is supported by the Gordon and Betty Moore Foundation (PI Kalogera, grant awards GBMF8477 and GBMF12341) and a Swiss National Science Foundation (PI Fragos, project numbers PP00P2\_211006 and CRSII5\_213497). J.J.A. acknowledges support for Program number (JWST-AR-04369.001-A) provided through a grant from the STScI under NASA contract NAS5-03127. The authors acknowledge UFIT Research Computing \url{http://www.rc.ufl.edu} for
providing computational resources and support that have contributed to the research results reported in this paper. This research was supported in part through the computational resources and staff contributions provided for the Quest high-performance computing facility at Northwestern University which is jointly supported by the Office of the Provost, the Office for Research, and Northwestern University Information Technology
\end{acknowledgments}

\begin{contribution}

A.C. led the writing and submission of the manuscript. J.J.A. coordinated individuals' effort and contributed to manuscript editing. J.J.A., T.F. and V.K. provided project supervision. M.B., S.G., P.M.S., and E.T. contributed to software development used in this work. All authors contributed to the reviewing of the manuscript.

\end{contribution}

\software{This manuscript has made use of the following Python modules: 
\texttt{numpy} \citep{2020NumPy-Array},
\texttt{scipy} \citep{scipy},
\texttt{pandas} \citep{pandas},
\texttt{matplotlib} \citep{matplotlib},
\texttt{astropy} 
\citep{astropy_I, astropy_II, astropy_III}
\texttt{scikit-learn} \citep{scikit-learn}. 
}

\appendix
\section{The Importance of Helium Star Winds}
\label{subsec:he_star_wind}

\begin{figure*}
    \centering
    \includegraphics[width=0.8\linewidth]{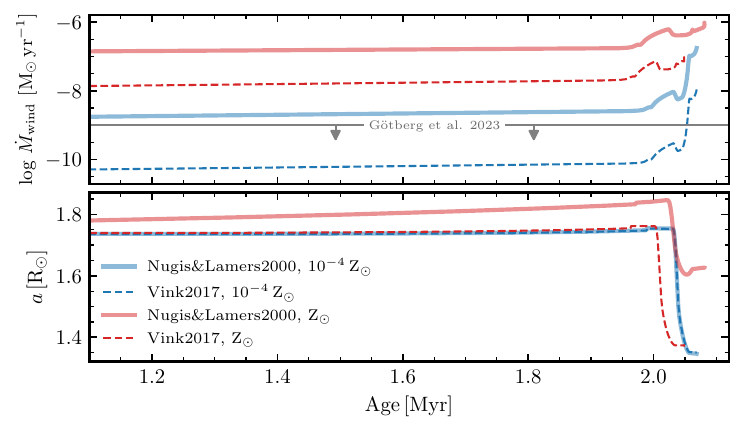}
    \caption{Comparison of helium star wind mass-loss prescriptions for a binary with a $\sim$ 2.9\Msun~He star and a 1.43\Msun~NS, with an initial orbital period of 0.13 days, at metallicities $Z=\Zsun$ (red) and $Z=10^{-4}\Zsun$ (blue). (\textit{Top row}) The wind mass-loss rates predicted by the \citet{Nugis&Lamers2000} and \citet{Vink2017} prescriptions differ by roughly one order-of magnitude at both metallicities, and exceed observational constraints from stripped stars in the Magellanic Clouds \citep[][horizontal gray line]{Gotberg+2023}. (\textit{Bottom row}) The evolution of the orbit leads to markedly different pre- and post-RLO separations, due to differing wind mass-loss rates under these prescriptions.}
    \label{fig:NL2000_vs_Vink2017}
\end{figure*}

In Section \ref{sec:physics_expectations}, we discuss how the single and binary processes prevalent during the He star$-$NS phase are responsible for setting the orbital scale at DNS birth. In this section, we discuss how our current limitations of the understanding of the He star wind mass losses affect the birth orbit size, especially at solar metallicity. Recent observations \citep{Gotberg+2023} suggest that \texttt{POSYDON}'s extrapolation of the \citet{Nugis&Lamers2000} WR prescription overstimates mass loss rates in the regime of low-mass He stars, which can lead to an observable difference in the orbital expansion (Section \ref{subsec:binary_rlo}). We therefore computed a set of \mesa models\footnote{The inlists are available at \href{https://doi.org/10.5281/zenodo.20516703}{doi:10.5281/zenodo.20516703}.} using the mass loss rates predicted from the Monte Carlo radiative transfer models in \citet{Vink2017}, since it is believed to be more accurate for low-mass He stars \citep[][and G\"otberg, Y., private communication]{Ramachandran+2024}. The resulting mass loss rates for the two prescriptions are shown in the top panel of Figure \ref{fig:NL2000_vs_Vink2017}. 

Although the two prescriptions differ by an order-of-magnitude, the mass loss rates at most metallicities remain higher than \citet{Gotberg+2023}, who find $\dot{M}\lesssim 10^{-9}\,M_\odot\,\rm{yr^{-1}}$ (gray horizontal line). The bottom panel of Figure \ref{fig:NL2000_vs_Vink2017} shows their impact on the orbital evolution of the system with $P_{\rm orb,\,i} = 0.13\rm{d}$ (from Fig. \ref{fig:orbital_evolution}). The choice of the wind prescription leads to markedly different pre- and post-RLO orbital separations, the contrast being more pronounced as we increase the metallicity. Adopting the weaker \citet{Vink2017} mass-loss rates reduces the orbital differences due to metallicity, which could misleadingly exaggerate its impact on the final DTD. As discussed in Section \ref{subsec:pop_synth_results}, prior evolution causes the binary orbits at the onset of the He star$-$NS phase to already vary with metallicity, making the effect of the He star winds only a minor perturbation.

From the two panels in Figure \ref{fig:NL2000_vs_Vink2017}, we infer that wind mass-loss rates $\lesssim 10^{-8}\,M_\odot\,\rm{yr^{-1}}$ will have no appreciable impact on the orbit. Guided by this insight, the \posydon binary population synthesis simulations used throughout this work were modified so that the evolution of the He star-NS phase at all metallicities (via the \texttt{CO-HeMS} grid) was artificially set to follow the evolutionary track of an equivalent binary in the $10^{-4}\,Z_\odot$ grid. While not exact, this ensures that the low wind mass-loss rates consistent with observations do not induce unphysical orbital expansion.

\bibliography{sample701}{}

\begin{thebibliography}{}
\expandafter\ifx\csname natexlab\endcsname\relax\def\natexlab#1{#1}\fi
\providecommand{\url}[1]{\href{#1}{#1}}
\providecommand{\dodoi}[1]{doi:~\href{http://doi.org/#1}{\nolinkurl{#1}}}
\providecommand{\doeprint}[1]{\href{http://ascl.net/#1}{\nolinkurl{http://ascl.net/#1}}}
\providecommand{\doarXiv}[1]{\href{https://arxiv.org/abs/#1}{\nolinkurl{https://arxiv.org/abs/#1}}}

% type= article
\bibitem[{B.~P. Abbott {et~al.}(2017{\natexlab{a}})Abbott
  {et~al.}}]{LIGOGW170817+2017}
Abbott, B.~P., {et~al.} 2017{\natexlab{a}}, \bibinfo{title}{{GW170817:
  Observation of Gravitational Waves from a Binary Neutron Star Inspiral},}
  Phys. Rev. Lett., 119, 161101, \dodoi{10.1103/PhysRevLett.119.161101}

% type= article
\bibitem[{B.~P. Abbott {et~al.}(2017{\natexlab{b}})Abbott
  {et~al.}}]{LIGOGW170817+kilonova}
Abbott, B.~P., {et~al.} 2017{\natexlab{b}}, \bibinfo{title}{{Multi-messenger
  Observations of a Binary Neutron Star Merger},} Astrophys. J. Lett., 848,
  L12, \dodoi{10.3847/2041-8213/aa91c9}

% type= article
\bibitem[{B.~P. Abbott {et~al.}(2020)Abbott {et~al.}}]{LIGOGW190425+2020}
Abbott, B.~P., {et~al.} 2020, \bibinfo{title}{{GW190425: Observation of a
  Compact Binary Coalescence with Total Mass $\sim 3.4 M_{\odot}$},} Astrophys.
  J. Lett., 892, L3, \dodoi{10.3847/2041-8213/ab75f5}

% type= article
\bibitem[{N. {Anand} {et~al.}(2018){Anand}, {Shahid}, \& {Resmi}}]{Anand+2018}
{Anand}, N., {Shahid}, M., \& {Resmi}, L. 2018, \bibinfo{title}{{Merger delay
  time distribution of extended emission short GRBs},} \mnras, 481, 4332,
  \dodoi{10.1093/mnras/sty2530}

% type= article
\bibitem[{J.~J. {Andrews} \& I. {Mandel}(2019){Andrews} \&
  {Mandel}}]{Andrews&Mandel2019}
{Andrews}, J.~J., \& {Mandel}, I. 2019, \bibinfo{title}{{Double Neutron Star
  Populations and Formation Channels},} \apjl, 880, L8,
  \dodoi{10.3847/2041-8213/ab2ed1}

% type= article
\bibitem[{J.~J. {Andrews} \& A. {Zezas}(2019){Andrews} \&
  {Zezas}}]{Andrews&Zezas2019}
{Andrews}, J.~J., \& {Zezas}, A. 2019, \bibinfo{title}{{Double neutron star
  formation: merger times, systemic velocities, and travel distances},} \mnras,
  486, 3213, \dodoi{10.1093/mnras/stz1066}

% type= article
\bibitem[{J.~J. {Andrews} {et~al.}(2025){Andrews}, {Bavera}, {Briel},
  {Chattaraj}, {Dotter}, {Fragos}, {Gallegos-Garcia}, {Gossage}, {Kalogera},
  {Kasdagli}, {Katsaggelos}, {Kimball}, {Kovlakas}, {Kruckow}, {Liotine},
  {Misra}, {Rocha}, {Souropanis}, {Srivastava}, {Sun}, {Teng}, {Xing},
  {Zapartas}, \& {Zevin}}]{Andrews+2025}
{Andrews}, J.~J., {Bavera}, S.~S., {Briel}, M., {et~al.} 2025,
  \bibinfo{title}{{POSYDON Version 2: Population Synthesis with Detailed
  Binary-evolution Simulations across a Cosmological Range of Metallicities},}
  \apjs, 281, 3, \dodoi{10.3847/1538-4365/adfb78}

% type= article
\bibitem[{ {Astropy Collaboration} {et~al.}(2013){Astropy Collaboration},
  {Robitaille}, {Tollerud}, {Greenfield}, {Droettboom}, {Bray}, {Aldcroft},
  {Davis}, {Ginsburg}, {Price-Whelan}, {Kerzendorf}, {Conley}, {Crighton},
  {Barbary}, {Muna}, {Ferguson}, {Grollier}, {Parikh}, {Nair}, {Unther},
  {Deil}, {Woillez}, {Conseil}, {Kramer}, {Turner}, {Singer}, {Fox}, {Weaver},
  {Zabalza}, {Edwards}, {Azalee Bostroem}, {Burke}, {Casey}, {Crawford},
  {Dencheva}, {Ely}, {Jenness}, {Labrie}, {Lim}, {Pierfederici}, {Pontzen},
  {Ptak}, {Refsdal}, {Servillat}, \& {Streicher}}]{astropy_I}
{Astropy Collaboration}, {Robitaille}, T.~P., {Tollerud}, E.~J., {et~al.} 2013,
  \bibinfo{title}{{Astropy: A community Python package for astronomy},} \aap,
  558, A33, \dodoi{10.1051/0004-6361/201322068}

% type= article
\bibitem[{ {Astropy Collaboration} {et~al.}(2018){Astropy Collaboration},
  {Price-Whelan}, {Sip{\H{o}}cz}, {G{\"u}nther}, {Lim}, {Crawford}, {Conseil},
  {Shupe}, {Craig}, {Dencheva}, {Ginsburg}, {VanderPlas}, {Bradley},
  {P{\'e}rez-Su{\'a}rez}, {de Val-Borro}, {Aldcroft}, {Cruz}, {Robitaille},
  {Tollerud}, {Ardelean}, {Babej}, {Bach}, {Bachetti}, {Bakanov}, {Bamford},
  {Barentsen}, {Barmby}, {Baumbach}, {Berry}, {Biscani}, {Boquien}, {Bostroem},
  {Bouma}, {Brammer}, {Bray}, {Breytenbach}, {Buddelmeijer}, {Burke},
  {Calderone}, {Cano Rodr{\'\i}guez}, {Cara}, {Cardoso}, {Cheedella}, {Copin},
  {Corrales}, {Crichton}, {D'Avella}, {Deil}, {Depagne}, {Dietrich}, {Donath},
  {Droettboom}, {Earl}, {Erben}, {Fabbro}, {Ferreira}, {Finethy}, {Fox},
  {Garrison}, {Gibbons}, {Goldstein}, {Gommers}, {Greco}, {Greenfield},
  {Groener}, {Grollier}, {Hagen}, {Hirst}, {Homeier}, {Horton}, {Hosseinzadeh},
  {Hu}, {Hunkeler}, {Ivezi{\'c}}, {Jain}, {Jenness}, {Kanarek}, {Kendrew},
  {Kern}, {Kerzendorf}, {Khvalko}, {King}, {Kirkby}, {Kulkarni}, {Kumar},
  {Lee}, {Lenz}, {Littlefair}, {Ma}, {Macleod}, {Mastropietro}, {McCully},
  {Montagnac}, {Morris}, {Mueller}, {Mumford}, {Muna}, {Murphy}, {Nelson},
  {Nguyen}, {Ninan}, {N{\"o}the}, {Ogaz}, {Oh}, {Parejko}, {Parley}, {Pascual},
  {Patil}, {Patil}, {Plunkett}, {Prochaska}, {Rastogi}, {Reddy Janga},
  {Sabater}, {Sakurikar}, {Seifert}, {Sherbert}, {Sherwood-Taylor}, {Shih},
  {Sick}, {Silbiger}, {Singanamalla}, {Singer}, {Sladen}, {Sooley},
  {Sornarajah}, {Streicher}, {Teuben}, {Thomas}, {Tremblay}, {Turner},
  {Terr{\'o}n}, {van Kerkwijk}, {de la Vega}, {Watkins}, {Weaver}, {Whitmore},
  {Woillez}, {Zabalza}, \& {Astropy Contributors}}]{astropy_II}
{Astropy Collaboration}, {Price-Whelan}, A.~M., {Sip{\H{o}}cz}, B.~M., {et~al.}
  2018, \bibinfo{title}{{The Astropy Project: Building an Open-science Project
  and Status of the v2.0 Core Package},} \aj, 156, 123,
  \dodoi{10.3847/1538-3881/aabc4f}

% type= article
\bibitem[{ {Astropy Collaboration} {et~al.}(2022){Astropy Collaboration},
  {Price-Whelan}, {Lim}, {Earl}, {Starkman}, {Bradley}, {Shupe}, {Patil},
  {Corrales}, {Brasseur}, {N{\"o}the}, {Donath}, {Tollerud}, {Morris},
  {Ginsburg}, {Vaher}, {Weaver}, {Tocknell}, {Jamieson}, {van Kerkwijk},
  {Robitaille}, {Merry}, {Bachetti}, {G{\"u}nther}, {Aldcroft},
  {Alvarado-Montes}, {Archibald}, {B{\'o}di}, {Bapat}, {Barentsen},
  {Baz{\'a}n}, {Biswas}, {Boquien}, {Burke}, {Cara}, {Cara}, {Conroy},
  {Conseil}, {Craig}, {Cross}, {Cruz}, {D'Eugenio}, {Dencheva}, {Devillepoix},
  {Dietrich}, {Eigenbrot}, {Erben}, {Ferreira}, {Foreman-Mackey}, {Fox},
  {Freij}, {Garg}, {Geda}, {Glattly}, {Gondhalekar}, {Gordon}, {Grant},
  {Greenfield}, {Groener}, {Guest}, {Gurovich}, {Handberg}, {Hart},
  {Hatfield-Dodds}, {Homeier}, {Hosseinzadeh}, {Jenness}, {Jones}, {Joseph},
  {Kalmbach}, {Karamehmetoglu}, {Ka{\l}uszy{\'n}ski}, {Kelley}, {Kern},
  {Kerzendorf}, {Koch}, {Kulumani}, {Lee}, {Ly}, {Ma}, {MacBride}, {Maljaars},
  {Muna}, {Murphy}, {Norman}, {O'Steen}, {Oman}, {Pacifici}, {Pascual},
  {Pascual-Granado}, {Patil}, {Perren}, {Pickering}, {Rastogi}, {Roulston},
  {Ryan}, {Rykoff}, {Sabater}, {Sakurikar}, {Salgado}, {Sanghi}, {Saunders},
  {Savchenko}, {Schwardt}, {Seifert-Eckert}, {Shih}, {Jain}, {Shukla}, {Sick},
  {Simpson}, {Singanamalla}, {Singer}, {Singhal}, {Sinha}, {Sip{\H{o}}cz},
  {Spitler}, {Stansby}, {Streicher}, {{\v{S}}umak}, {Swinbank}, {Taranu},
  {Tewary}, {Tremblay}, {de Val-Borro}, {Van Kooten}, {Vasovi{\'c}}, {Verma},
  {de Miranda Cardoso}, {Williams}, {Wilson}, {Winkel}, {Wood-Vasey}, {Xue},
  {Yoachim}, {Zhang}, {Zonca}, \& {Astropy Project Contributors}}]{astropy_III}
{Astropy Collaboration}, {Price-Whelan}, A.~M., {Lim}, P.~L., {et~al.} 2022,
  \bibinfo{title}{{The Astropy Project: Sustaining and Growing a
  Community-oriented Open-source Project and the Latest Major Release (v5.0) of
  the Core Package},} \apj, 935, 167, \dodoi{10.3847/1538-4357/ac7c74}

% type= article
\bibitem[{A. {Bandyopadhyay} {et~al.}(2024){Bandyopadhyay}, {Ezzeddine},
  {Allende Prieto}, {Aria}, {Shah}, {Beers}, {Frebel}, {Hansen}, {Holmbeck},
  {Placco}, {Roederer}, \& {Sakari}}]{Bandyopadhyay+2024}
{Bandyopadhyay}, A., {Ezzeddine}, R., {Allende Prieto}, C., {et~al.} 2024,
  \bibinfo{title}{{The R-process Alliance: Fifth Data Release from the Search
  for R-process-enhanced Metal-poor Stars in the Galactic Halo with the GTC},}
  \apjs, 274, 39, \dodoi{10.3847/1538-4365/ad6f0f}

% type= article
\bibitem[{P.~S. {Behroozi} {et~al.}(2014){Behroozi}, {Ramirez-Ruiz}, \&
  {Fryer}}]{Behroozi+2014}
{Behroozi}, P.~S., {Ramirez-Ruiz}, E., \& {Fryer}, C.~L. 2014,
  \bibinfo{title}{{Interpreting Short Gamma-Ray Burst Progenitor Kicks and Time
  Delays using the Host Galaxy-Dark Matter Halo Connection},} \apj, 792, 123,
  \dodoi{10.1088/0004-637X/792/2/123}

% type= article
\bibitem[{K. {Belczynski} {et~al.}(2002){Belczynski}, {Kalogera}, \&
  {Bulik}}]{Belczynski+2002}
{Belczynski}, K., {Kalogera}, V., \& {Bulik}, T. 2002, \bibinfo{title}{{A
  Comprehensive Study of Binary Compact Objects as Gravitational Wave Sources:
  Evolutionary Channels, Rates, and Physical Properties},} \apj, 572, 407,
  \dodoi{10.1086/340304}

% type= article
\bibitem[{K. {Belczynski} {et~al.}(2006){Belczynski}, {Perna}, {Bulik},
  {Kalogera}, {Ivanova}, \& {Lamb}}]{Belczynski+2006}
{Belczynski}, K., {Perna}, R., {Bulik}, T., {et~al.} 2006, \bibinfo{title}{{A
  Study of Compact Object Mergers as Short Gamma-Ray Burst Progenitors},} \apj,
  648, 1110, \dodoi{10.1086/505169}

% type= article
\bibitem[{K. {Belczynski} {et~al.}(2018{\natexlab{a}}){Belczynski}, {Askar},
  {Arca-Sedda}, {Chruslinska}, {Donnari}, {Giersz}, {Benacquista}, {Spurzem},
  {Jin}, {Wiktorowicz}, \& {Belloni}}]{Belczynski+2018_DNSmerger}
{Belczynski}, K., {Askar}, A., {Arca-Sedda}, M., {et~al.} 2018{\natexlab{a}},
  \bibinfo{title}{{The origin of the first neutron star - neutron star
  merger},} \aap, 615, A91, \dodoi{10.1051/0004-6361/201732428}

% type= article
\bibitem[{K. {Belczynski} {et~al.}(2018{\natexlab{b}}){Belczynski}, {Bulik},
  {Olejak}, {Chruslinska}, {Singh}, {Pol}, {Zdunik}, {O'Shaughnessy},
  {McLaughlin}, {Lorimer}, {Korobkin}, {van den Heuvel}, {Davies}, \&
  {Holz}}]{Belczynski+2018}
{Belczynski}, K., {Bulik}, T., {Olejak}, A., {et~al.} 2018{\natexlab{b}},
  \bibinfo{title}{{Binary neutron star formation and the origin of GW170817},}
  arXiv e-prints, arXiv:1812.10065, \dodoi{10.48550/arXiv.1812.10065}

% type= article
\bibitem[{P. {Beniamini} {et~al.}(2016){Beniamini}, {Hotokezaka}, \&
  {Piran}}]{Beniamini+2016}
{Beniamini}, P., {Hotokezaka}, K., \& {Piran}, T. 2016, \bibinfo{title}{{Natal
  Kicks and Time Delays in Merging Neutron Star Binaries: Implications for
  r-process Nucleosynthesis in Ultra-faint Dwarfs and in the Milky Way},}
  \apjl, 829, L13, \dodoi{10.3847/2041-8205/829/1/L13}

% type= article
\bibitem[{P. {Beniamini} \& T. {Piran}(2016){Beniamini} \&
  {Piran}}]{Beniamini&Piran2016}
{Beniamini}, P., \& {Piran}, T. 2016, \bibinfo{title}{{Formation of double
  neutron star systems as implied by observations},} \mnras, 456, 4089,
  \dodoi{10.1093/mnras/stv2903}

% type= article
\bibitem[{P. {Beniamini} \& T. {Piran}(2019){Beniamini} \&
  {Piran}}]{Beniamini&Piran2019}
{Beniamini}, P., \& {Piran}, T. 2019, \bibinfo{title}{{The Gravitational waves
  merger time distribution of binary neutron star systems},} \mnras, 487, 4847,
  \dodoi{10.1093/mnras/stz1589}

% type= article
\bibitem[{P. {Beniamini} \& T. {Piran}(2024){Beniamini} \&
  {Piran}}]{Beniamini&Piran2024}
{Beniamini}, P., \& {Piran}, T. 2024, \bibinfo{title}{{Ultrafast Compact Binary
  Mergers},} \apj, 966, 17, \dodoi{10.3847/1538-4357/ad32cd}

% type= article
\bibitem[{E. {Berger} {et~al.}(2007){Berger}, {Fox}, {Price}, {Nakar},
  {Gal-Yam}, {Holz}, {Schmidt}, {Cucchiara}, {Cenko}, {Kulkarni}, {Soderberg},
  {Frail}, {Penprase}, {Rau}, {Ofek}, {Burnell}, {Cameron}, {Cowie}, {Dopita},
  {Hook}, {Peterson}, {Podsiadlowski}, {Roth}, {Rutledge}, {Sheppard}, \&
  {Songaila}}]{Berger+2007}
{Berger}, E., {Fox}, D.~B., {Price}, P.~A., {et~al.} 2007, \bibinfo{title}{{A
  New Population of High-Redshift Short-Duration Gamma-Ray Bursts},} \apj, 664,
  1000, \dodoi{10.1086/518762}

% type= article
\bibitem[{A. {Blaauw}(1961){Blaauw}}]{Blaauw1961}
{Blaauw}, A. 1961, \bibinfo{title}{{On the origin of the O- and B-type stars
  with high velocities (the ``run-away'' stars), and some related problems},}
  \bain, 15, 265

% type= article
\bibitem[{M. {Bonetti} {et~al.}(2019){Bonetti}, {Perego}, {Dotti}, \&
  {Cescutti}}]{Bonetti+2019}
{Bonetti}, M., {Perego}, A., {Dotti}, M., \& {Cescutti}, G. 2019,
  \bibinfo{title}{{Neutron star binary orbits in their host potential: effect
  on early r-process enrichment},} \mnras, 490, 296,
  \dodoi{10.1093/mnras/stz2554}

% type= article
\bibitem[{F.~S. {Broekgaarden} {et~al.}(2022){Broekgaarden}, {Berger},
  {Stevenson}, {Justham}, {Mandel}, {Chru{\'s}li{\'n}ska}, {van Son}, {Wagg},
  {Vigna-G{\'o}mez}, {de Mink}, {Chattopadhyay}, \&
  {Neijssel}}]{Broekgaarden+2022}
{Broekgaarden}, F.~S., {Berger}, E., {Stevenson}, S., {et~al.} 2022,
  \bibinfo{title}{{Impact of massive binary star and cosmic evolution on
  gravitational wave observations - II. Double compact object rates and
  properties},} \mnras, 516, 5737, \dodoi{10.1093/mnras/stac1677}

% type= article
\bibitem[{M. {Cantiello} {et~al.}(2009){Cantiello}, {Langer}, {Brott}, {de
  Koter}, {Shore}, {Vink}, {Voegler}, {Lennon}, \& {Yoon}}]{Cantiello+2009}
{Cantiello}, M., {Langer}, N., {Brott}, I., {et~al.} 2009,
  \bibinfo{title}{{Sub-surface convection zones in hot massive stars and their
  observable consequences},} \aap, 499, 279,
  \dodoi{10.1051/0004-6361/200911643}

% type= article
\bibitem[{J.~I. {Castor} {et~al.}(1975){Castor}, {Abbott}, \&
  {Klein}}]{Castor+1975}
{Castor}, J.~I., {Abbott}, D.~C., \& {Klein}, R.~I. 1975,
  \bibinfo{title}{{Radiation-driven winds in Of stars.},} \apj, 195, 157,
  \dodoi{10.1086/153315}

% type= article
\bibitem[{J. {Cehula} \& O. {Pejcha}(2023){Cehula} \&
  {Pejcha}}]{Cehula&Pejcha2023}
{Cehula}, J., \& {Pejcha}, O. 2023, \bibinfo{title}{{A theory of mass transfer
  in binary stars},} \mnras, 524, 471, \dodoi{10.1093/mnras/stad1862}

% type= article
\bibitem[{A. {Chattaraj} {et~al.}(2026){Chattaraj}, {Andrews}, {Bavera},
  {Briel}, {Chattopadhyay}, {Fragos}, {Gossage}, {Kalogera}, {Kovlakas},
  {Kruckow}, {Liotine}, {Rocha}, {Srivastava}, {Sun}, {Teng}, {Xing}, \&
  {Zapartas}}]{Chattaraj+2026}
{Chattaraj}, A., {Andrews}, J.~J., {Bavera}, S.~S., {et~al.} 2026,
  \bibinfo{title}{{Forming Double Neutron Stars Using Detailed Binary Evolution
  Models with POSYDON: Comparison to the Galactic Systems},} \apj, 997, 52,
  \dodoi{10.3847/1538-4357/ae1f93}

% type= article
\bibitem[{A. Chattaraj {et~al.}(in prep.)Chattaraj
  {et~al.}}]{Chattaraj+2026_in-prep}
Chattaraj, A., {et~al.} in prep.

% type= article
\bibitem[{H.-Y. {Chen} {et~al.}(2025){Chen}, {Landry}, {Read}, \&
  {Siegel}}]{Chen+2025}
{Chen}, H.-Y., {Landry}, P., {Read}, J.~S., \& {Siegel}, D.~M. 2025,
  \bibinfo{title}{{Inference of Multichannel r-process Element Enrichment in
  the Milky Way Using Binary Neutron Star Merger Observations},} \apj, 985,
  154, \dodoi{10.3847/1538-4357/add0af}

% type= article
\bibitem[{M. {Chruslinska} {et~al.}(2018){Chruslinska}, {Belczynski},
  {Klencki}, \& {Benacquista}}]{Chruslinska+2018}
{Chruslinska}, M., {Belczynski}, K., {Klencki}, J., \& {Benacquista}, M. 2018,
  \bibinfo{title}{{Double neutron stars: merger rates revisited},} \mnras, 474,
  2937, \dodoi{10.1093/mnras/stx2923}

% type= article
\bibitem[{Q. {Chu} {et~al.}(2022){Chu}, {Yu}, \& {Lu}}]{Chu+2022}
{Chu}, Q., {Yu}, S., \& {Lu}, Y. 2022, \bibinfo{title}{{Formation and evolution
  of binary neutron stars: mergers and their host galaxies},} \mnras, 509,
  1557, \dodoi{10.1093/mnras/stab2882}

% type= article
\bibitem[{B. {C{\^o}t{\'e}} {et~al.}(2019){C{\^o}t{\'e}}, {Eichler}, {Arcones},
  {Hansen}, {Simonetti}, {Frebel}, {Fryer}, {Pignatari}, {Reichert},
  {Belczynski}, \& {Matteucci}}]{Cote+2019}
{C{\^o}t{\'e}}, B., {Eichler}, M., {Arcones}, A., {et~al.} 2019,
  \bibinfo{title}{{Neutron Star Mergers Might Not Be the Only Source of
  r-process Elements in the Milky Way},} \apj, 875, 106,
  \dodoi{10.3847/1538-4357/ab10db}

% type= article
\bibitem[{A.~N. {Cox} \& J.~E. {Tabor}(1976){Cox} \& {Tabor}}]{Cox&Tabor1976}
{Cox}, A.~N., \& {Tabor}, J.~E. 1976, \bibinfo{title}{{Radiative opacity tables
  for 40 stellar mixtures.},} \apjs, 31, 271, \dodoi{10.1086/190383}

% type= article
\bibitem[{A.~L. {De Santis} {et~al.}(2026){De Santis}, {Ronchini},
  {Santoliquido}, \& {Branchesi}}]{DeSantis+2026}
{De Santis}, A.~L., {Ronchini}, S., {Santoliquido}, F., \& {Branchesi}, M.
  2026, \bibinfo{title}{{Constraining Binary Neutron Star Populations using
  Short Gamma-Ray Burst Observations},} arXiv e-prints, arXiv:2602.13391,
  \dodoi{10.48550/arXiv.2602.13391}

% type= article
\bibitem[{A.~J. {Delgado} \& H.~C. {Thomas}(1981){Delgado} \&
  {Thomas}}]{Delgado&Thomas1981}
{Delgado}, A.~J., \& {Thomas}, H.~C. 1981, \bibinfo{title}{{Mass transfer in a
  binary system - The evolution of the mass-giving helium star},} \aap, 96, 142

% type= article
\bibitem[{J.~D.~M. {Dewi} \& O.~R. {Pols}(2003){Dewi} \&
  {Pols}}]{Dewi&Pols2003}
{Dewi}, J.~D.~M., \& {Pols}, O.~R. 2003, \bibinfo{title}{{The late stages of
  evolution of helium star-neutron star binaries and the formation of double
  neutron star systems},} \mnras, 344, 629,
  \dodoi{10.1046/j.1365-8711.2003.06844.x}

% type= article
\bibitem[{J.~D.~M. {Dewi} {et~al.}(2002){Dewi}, {Pols}, {Savonije}, \& {van den
  Heuvel}}]{Dewi+2002}
{Dewi}, J.~D.~M., {Pols}, O.~R., {Savonije}, G.~J., \& {van den Heuvel},
  E.~P.~J. 2002, \bibinfo{title}{{The evolution of naked helium stars with a
  neutron star companion in close binary systems},} \mnras, 331, 1027,
  \dodoi{10.1046/j.1365-8711.2002.05257.x}

% type= article
\bibitem[{P. {Disberg} {et~al.}(2024){Disberg}, {Gaspari}, \&
  {Levan}}]{Disberg+2024}
{Disberg}, P., {Gaspari}, N., \& {Levan}, A.~J. 2024,
  \bibinfo{title}{{Kinematic constraints on the ages and kick velocities of
  Galactic neutron star binaries},} \aap, 689, A348,
  \dodoi{10.1051/0004-6361/202450790}

% type= article
\bibitem[{P. {Disberg} \& I. {Mandel}(2025){Disberg} \&
  {Mandel}}]{Disberg&Mandel2025}
{Disberg}, P., \& {Mandel}, I. 2025, \bibinfo{title}{{The Kick Velocity
  Distribution of Isolated Neutron Stars},} arXiv e-prints, arXiv:2505.22102,
  \dodoi{10.48550/arXiv.2505.22102}

% type= article
\bibitem[{M. {Dominik} {et~al.}(2012){Dominik}, {Belczynski}, {Fryer}, {Holz},
  {Berti}, {Bulik}, {Mandel}, \& {O'Shaughnessy}}]{Dominik+2012}
{Dominik}, M., {Belczynski}, K., {Fryer}, C., {et~al.} 2012,
  \bibinfo{title}{{Double Compact Objects. I. The Significance of the Common
  Envelope on Merger Rates},} \apj, 759, 52, \dodoi{10.1088/0004-637X/759/1/52}

% type= article
\bibitem[{P.~P. {Eggleton}(1983){Eggleton}}]{Eggleton1983}
{Eggleton}, P.~P. 1983, \bibinfo{title}{{Aproximations to the radii of Roche
  lobes.},} \apj, 268, 368, \dodoi{10.1086/160960}

% type= article
\bibitem[{M. {Fishbach} {et~al.}(2026){Fishbach}, {Ji}, {Fong}, {Wu},
  {Rastinejad}, {Vijaykumar}, \& {Chen}}]{Fishbach+2026}
{Fishbach}, M., {Ji}, A.~P., {Fong}, W.-f., {et~al.} 2026,
  \bibinfo{title}{{Implications of low neutron star merger rates for gamma-ray
  bursts, r-process production and Galactic double neutron stars},} arXiv
  e-prints, arXiv:2604.05059.
\newblock \doarXiv{2604.05059}

% type= article
\bibitem[{B.~P. {Flannery} \& E.~P.~J. {van den Heuvel}(1975){Flannery} \& {van
  den Heuvel}}]{Flannery&vandenHeuvel1975}
{Flannery}, B.~P., \& {van den Heuvel}, E.~P.~J. 1975, \bibinfo{title}{{On the
  origin of the binary pulsar PSR 1913+16.},} \aap, 39, 61

% type= article
\bibitem[{W. {Fong} {et~al.}(2013){Fong}, {Berger}, {Chornock}, {Margutti},
  {Levan}, {Tanvir}, {Tunnicliffe}, {Czekala}, {Fox}, {Perley}, {Cenko},
  {Zauderer}, {Laskar}, {Persson}, {Monson}, {Kelson}, {Birk}, {Murphy},
  {Servillat}, \& {Anglada}}]{Fong+2013}
{Fong}, W., {Berger}, E., {Chornock}, R., {et~al.} 2013,
  \bibinfo{title}{{Demographics of the Galaxies Hosting Short-duration
  Gamma-Ray Bursts},} \apj, 769, 56, \dodoi{10.1088/0004-637X/769/1/56}

% type= article
\bibitem[{T. {Fragos} {et~al.}(2023){Fragos}, {Andrews}, {Bavera}, {Berry},
  {Coughlin}, {Dotter}, {Giri}, {Kalogera}, {Katsaggelos}, {Kovlakas},
  {Lalvani}, {Misra}, {Srivastava}, {Qin}, {Rocha}, {Rom{\'a}n-Garza}, {Serra},
  {Stahle}, {Sun}, {Teng}, {Trajcevski}, {Tran}, {Xing}, {Zapartas}, \&
  {Zevin}}]{Fragos+2023}
{Fragos}, T., {Andrews}, J.~J., {Bavera}, S.~S., {et~al.} 2023,
  \bibinfo{title}{{POSYDON: A General-purpose Population Synthesis Code with
  Detailed Binary-evolution Simulations},} \apjs, 264, 45,
  \dodoi{10.3847/1538-4365/ac90c1}

% type= article
\bibitem[{M. {Gallegos-Garcia} {et~al.}(2023){Gallegos-Garcia}, {Berry}, \&
  {Kalogera}}]{Gallegos-Garcia+2023}
{Gallegos-Garcia}, M., {Berry}, C. P.~L., \& {Kalogera}, V. 2023,
  \bibinfo{title}{{Evolutionary Origins of Binary Neutron Star Mergers: Effects
  of Common Envelope Efficiency and Metallicity},} \apj, 955, 133,
  \dodoi{10.3847/1538-4357/ace434}

% type= article
\bibitem[{Y. {G{\"o}tberg} {et~al.}(2017){G{\"o}tberg}, {de Mink}, \&
  {Groh}}]{Gotberg+2017}
{G{\"o}tberg}, Y., {de Mink}, S.~E., \& {Groh}, J.~H. 2017,
  \bibinfo{title}{{Ionizing spectra of stars that lose their envelope through
  interaction with a binary companion: role of metallicity},} \aap, 608, A11,
  \dodoi{10.1051/0004-6361/201730472}

% type= article
\bibitem[{Y. {G{\"o}tberg} {et~al.}(2023){G{\"o}tberg}, {Drout}, {Ji}, {Groh},
  {Ludwig}, {Crowther}, {Smith}, {de Koter}, \& {de Mink}}]{Gotberg+2023}
{G{\"o}tberg}, Y., {Drout}, M.~R., {Ji}, A.~P., {et~al.} 2023,
  \bibinfo{title}{{Stellar Properties of Observed Stars Stripped in Binaries in
  the Magellanic Clouds},} \apj, 959, 125, \dodoi{10.3847/1538-4357/ace5a3}

% type= article
\bibitem[{Y.-L. {Guo} {et~al.}(2024){Guo}, {Wang}, {Chen}, {Li}, {Ge}, {Jiang},
  \& {Han}}]{Guo+2024}
{Guo}, Y.-L., {Wang}, B., {Chen}, W.-C., {et~al.} 2024,
  \bibinfo{title}{{Electron-capture supernovae in NS + He star systems and the
  double neutron star systems},} \mnras, 530, 4461,
  \dodoi{10.1093/mnras/stae1112}

% type= article
\bibitem[{T.~T. {Hansen} {et~al.}(2017){Hansen}, {Simon}, {Marshall}, {Li},
  {Carollo}, {DePoy}, {Nagasawa}, {Bernstein}, {Drlica-Wagner}, {Abdalla},
  {Allam}, {Annis}, {Bechtol}, {Benoit-L{\'e}vy}, {Brooks}, {Buckley-Geer},
  {Carnero Rosell}, {Carrasco Kind}, {Carretero}, {Cunha}, {da Costa}, {Desai},
  {Eifler}, {Fausti Neto}, {Flaugher}, {Frieman}, {Garc{\'\i}a-Bellido},
  {Gaztanaga}, {Gerdes}, {Gruen}, {Gruendl}, {Gschwend}, {Gutierrez}, {James},
  {Krause}, {Kuehn}, {Kuropatkin}, {Lahav}, {Miquel}, {Plazas}, {Romer},
  {Sanchez}, {Santiago}, {Scarpine}, {Smith}, {Soares-Santos}, {Sobreira},
  {Suchyta}, {Swanson}, {Tarle}, {Walker}, \& {DES
  Collaboration}}]{Hansen+2017}
{Hansen}, T.~T., {Simon}, J.~D., {Marshall}, J.~L., {et~al.} 2017,
  \bibinfo{title}{{An r-process Enhanced Star in the Dwarf Galaxy Tucana III},}
  \apj, 838, 44, \dodoi{10.3847/1538-4357/aa634a}

% type= article
\bibitem[{J.-M. {Hao} \& Y.-F. {Yuan}(2013){Hao} \& {Yuan}}]{Hao&Yuan2013}
{Hao}, J.-M., \& {Yuan}, Y.-F. 2013, \bibinfo{title}{{Progenitor delay-time
  distribution of short gamma-ray bursts: Constraints from observations},}
  \aap, 558, A22, \dodoi{10.1051/0004-6361/201321471}

% type= article
\bibitem[{C.~R. Harris {et~al.}(2020)Harris, Millman, van~der Walt, Gommers,
  Virtanen, Cournapeau, Wieser, Taylor, Berg, Smith, Kern, Picus, Hoyer, van
  Kerkwijk, Brett, Haldane, Fernández~del Río, Wiebe, Peterson,
  Gérard-Marchant, Sheppard, Reddy, Weckesser, Abbasi, Gohlke, \&
  Oliphant}]{2020NumPy-Array}
Harris, C.~R., Millman, K.~J., van~der Walt, S.~J., {et~al.} 2020,
  \bibinfo{title}{Array programming with {NumPy},} Nature, 585, 357–362,
  \dodoi{10.1038/s41586-020-2649-2}

% type= article
\bibitem[{L.~E. {Henderson} {et~al.}(2025{\natexlab{a}}){Henderson},
  {Gerasimov}, \& {Kirby}}]{Henderson+2025b}
{Henderson}, L.~E., {Gerasimov}, R., \& {Kirby}, E.~N. 2025{\natexlab{a}},
  \bibinfo{title}{{Population-dependent r-process Scatter in the Globular
  Cluster M15},} \apjl, 992, L14, \dodoi{10.3847/2041-8213/ae0a4a}

% type= article
\bibitem[{L.~E. {Henderson} {et~al.}(2025{\natexlab{b}}){Henderson}, {Kirby},
  {de los Reyes}, {Gerasimov}, \& {Manwadkar}}]{Henderson+2025a}
{Henderson}, L.~E., {Kirby}, E.~N., {de los Reyes}, M. A.~C., {Gerasimov}, R.,
  \& {Manwadkar}, V. 2025{\natexlab{b}}, \bibinfo{title}{{Neutron-capture
  Element Abundances of 491 Stars in Milky Way Dwarf Satellite Galaxies from
  Medium-resolution Spectra},} \apj, 983, 117, \dodoi{10.3847/1538-4357/adbe7d}

% type= article
\bibitem[{V. {Hill} {et~al.}(2002){Hill}, {Plez}, {Cayrel}, {Beers},
  {Nordstr{\"o}m}, {Andersen}, {Spite}, {Spite}, {Barbuy}, {Bonifacio},
  {Depagne}, {Fran{\c{c}}ois}, \& {Primas}}]{Hill+2002}
{Hill}, V., {Plez}, B., {Cayrel}, R., {et~al.} 2002, \bibinfo{title}{{First
  stars. I. The extreme r-element rich, iron-poor halo giant CS 31082-001.
  Implications for the r-process site(s) and radioactive cosmochronology},}
  \aap, 387, 560, \dodoi{10.1051/0004-6361:20020434}

% type= article
\bibitem[{G. {Hobbs} {et~al.}(2005){Hobbs}, {Lorimer}, {Lyne}, \&
  {Kramer}}]{Hobbs+2005}
{Hobbs}, G., {Lorimer}, D.~R., {Lyne}, A.~G., \& {Kramer}, M. 2005,
  \bibinfo{title}{{A statistical study of 233 pulsar proper motions},} \mnras,
  360, 974, \dodoi{10.1111/j.1365-2966.2005.09087.x}

% type= article
\bibitem[{E.~M. {Holmbeck} \& J.~J. {Andrews}(2024){Holmbeck} \&
  {Andrews}}]{Holmbeck&Andrews2024}
{Holmbeck}, E.~M., \& {Andrews}, J.~J. 2024, \bibinfo{title}{{Total r-process
  Yields of Milky Way Neutron Star Mergers},} \apj, 963, 110,
  \dodoi{10.3847/1538-4357/ad1e52}

% type= article
\bibitem[{K. {Hotokezaka} {et~al.}(2018){Hotokezaka}, {Beniamini}, \&
  {Piran}}]{Hotokezaka+2018}
{Hotokezaka}, K., {Beniamini}, P., \& {Piran}, T. 2018,
  \bibinfo{title}{{Neutron star mergers as sites of r-process nucleosynthesis
  and short gamma-ray bursts},} International Journal of Modern Physics D, 27,
  1842005, \dodoi{10.1142/S0218271818420051}

% type= article
\bibitem[{J.~D. Hunter(2007)Hunter}]{matplotlib}
Hunter, J.~D. 2007, \bibinfo{title}{Matplotlib: A 2D graphics environment,}
  Computing in Science \& Engineering, 9, 90, \dodoi{10.1109/MCSE.2007.55}

% type= article
\bibitem[{C.~A. {Iglesias} \& F.~J. {Rogers}(1991){Iglesias} \&
  {Rogers}}]{Iglesias&Rogers1991}
{Iglesias}, C.~A., \& {Rogers}, F.~J. 1991, \bibinfo{title}{{Opacities for the
  Solar Radiative Interior},} \apj, 371, 408, \dodoi{10.1086/169902}

% type= article
\bibitem[{C.~A. {Iglesias} {et~al.}(1992){Iglesias}, {Rogers}, \&
  {Wilson}}]{Iglesias+1992}
{Iglesias}, C.~A., {Rogers}, F.~J., \& {Wilson}, B.~G. 1992,
  \bibinfo{title}{{Spin-Orbit Interaction Effects on the Rosseland Mean
  Opacity},} \apj, 397, 717, \dodoi{10.1086/171827}

% type= article
\bibitem[{M. {Im} {et~al.}(2017){Im}, {Yoon}, {Lee}, {Lee}, {Kim}, {Lee},
  {Kim}, {Troja}, {Choi}, {Lim}, {Ko}, \& {Shim}}]{Im+2017}
{Im}, M., {Yoon}, Y., {Lee}, S.-K.~J., {et~al.} 2017, \bibinfo{title}{{Distance
  and Properties of NGC 4993 as the Host Galaxy of the Gravitational-wave
  Source GW170817},} \apjl, 849, L16, \dodoi{10.3847/2041-8213/aa9367}

% type= article
\bibitem[{N. {Ivanova} {et~al.}(2003){Ivanova}, {Belczynski}, {Kalogera},
  {Rasio}, \& {Taam}}]{Ivanova+2003}
{Ivanova}, N., {Belczynski}, K., {Kalogera}, V., {Rasio}, F.~A., \& {Taam},
  R.~E. 2003, \bibinfo{title}{{The Role of Helium Stars in the Formation of
  Double Neutron Stars},} \apj, 592, 475, \dodoi{10.1086/375578}

% type= article
\bibitem[{H.-T. {Janka}(2017){Janka}}]{Janka2017}
{Janka}, H.-T. 2017, \bibinfo{title}{{Neutron Star Kicks by the Gravitational
  Tug-boat Mechanism in Asymmetric Supernova Explosions: Progenitor and
  Explosion Dependence},} \apj, 837, 84, \dodoi{10.3847/1538-4357/aa618e}

% type= article
\bibitem[{A.~S. {Jermyn} {et~al.}(2023){Jermyn}, {Bauer}, {Schwab}, {Farmer},
  {Ball}, {Bellinger}, {Dotter}, {Joyce}, {Marchant}, {Mombarg}, {Wolf}, {Sunny
  Wong}, {Cinquegrana}, {Farrell}, {Smolec}, {Thoul}, {Cantiello}, {Herwig},
  {Toloza}, {Bildsten}, {Townsend}, \& {Timmes}}]{Jermyn+2023}
{Jermyn}, A.~S., {Bauer}, E.~B., {Schwab}, J., {et~al.} 2023,
  \bibinfo{title}{{Modules for Experiments in Stellar Astrophysics (MESA):
  Time-dependent Convection, Energy Conservation, Automatic Differentiation,
  and Infrastructure},} \apjs, 265, 15, \dodoi{10.3847/1538-4365/acae8d}

% type= article
\bibitem[{A.~P. {Ji} {et~al.}(2016{\natexlab{a}}){Ji}, {Frebel}, {Chiti}, \&
  {Simon}}]{Ji+2016_Nature}
{Ji}, A.~P., {Frebel}, A., {Chiti}, A., \& {Simon}, J.~D. 2016{\natexlab{a}},
  \bibinfo{title}{{R-process enrichment from a single event in an ancient dwarf
  galaxy},} \nat, 531, 610, \dodoi{10.1038/nature17425}

% type= article
\bibitem[{A.~P. {Ji} {et~al.}(2016{\natexlab{b}}){Ji}, {Frebel}, {Simon}, \&
  {Chiti}}]{Ji+2016}
{Ji}, A.~P., {Frebel}, A., {Simon}, J.~D., \& {Chiti}, A. 2016{\natexlab{b}},
  \bibinfo{title}{{Complete Element Abundances of Nine Stars in the r-process
  Galaxy Reticulum II},} \apj, 830, 93, \dodoi{10.3847/0004-637X/830/2/93}

% type= article
\bibitem[{A.~P. {Ji} {et~al.}(2023){Ji}, {Simon}, {Roederer}, {Magg}, {Frebel},
  {Johnson}, {Klessen}, {Magg}, {Cescutti}, {Mateo}, {Bergemann}, \&
  {Bailey}}]{Ji+2023}
{Ji}, A.~P., {Simon}, J.~D., {Roederer}, I.~U., {et~al.} 2023,
  \bibinfo{title}{{Metal Mixing in the r-process Enhanced Ultrafaint Dwarf
  Galaxy Reticulum II},} \aj, 165, 100, \dodoi{10.3847/1538-3881/acad84}

% type= article
\bibitem[{L. {Jiang} {et~al.}(2024){Jiang}, {Xu}, {Zha}, {Guo}, {Yuan}, {Qian},
  {Chen}, \& {Wang}}]{Jiang+2024}
{Jiang}, L., {Xu}, K., {Zha}, S., {et~al.} 2024, \bibinfo{title}{{On the
  Formation of the Double Neutron Star Binary PSR J1846{\textendash}0513},}
  Research in Astronomy and Astrophysics, 24, 115022,
  \dodoi{10.1088/1674-4527/ad8d1b}

% type= article
\bibitem[{V. {Kalogera}(1996){Kalogera}}]{Kaolgera1996}
{Kalogera}, V. 1996, \bibinfo{title}{{Orbital Characteristics of Binary Systems
  after Asymmetric Supernova Explosions},} \apj, 471, 352,
  \dodoi{10.1086/177974}

% type= article
\bibitem[{D. {Kasen} {et~al.}(2017){Kasen}, {Metzger}, {Barnes}, {Quataert}, \&
  {Ramirez-Ruiz}}]{Kasen+2017}
{Kasen}, D., {Metzger}, B., {Barnes}, J., {Quataert}, E., \& {Ramirez-Ruiz}, E.
  2017, \bibinfo{title}{{Origin of the heavy elements in binary neutron-star
  mergers from a gravitational-wave event},} \nat, 551, 80,
  \dodoi{10.1038/nature24453}

% type= book
\bibitem[{R. {Kippenhahn} \& A. {Weigert}(1994){Kippenhahn} \&
  {Weigert}}]{Kippenhahn&Weigert1994}
{Kippenhahn}, R., \& {Weigert}, A. 1994, {Stellar Structure and Evolution}

% type= article
\bibitem[{E.~N. {Kirby} {et~al.}(2023){Kirby}, {Ji}, \& {Kovalev}}]{Kirby+2023}
{Kirby}, E.~N., {Ji}, A.~P., \& {Kovalev}, M. 2023, \bibinfo{title}{{r-process
  Abundance Patterns in the Globular Cluster M92},} \apj, 958, 45,
  \dodoi{10.3847/1538-4357/acf309}

% type= article
\bibitem[{C. {Kobayashi} {et~al.}(2023){Kobayashi}, {Mandel}, {Belczynski},
  {Goriely}, {Janka}, {Just}, {Ruiter}, {Vanbeveren}, {Kruckow}, {Briel},
  {Eldridge}, \& {Stanway}}]{Kobayashi+2023}
{Kobayashi}, C., {Mandel}, I., {Belczynski}, K., {et~al.} 2023,
  \bibinfo{title}{{Can Neutron Star Mergers Alone Explain the r-process
  Enrichment of the Milky Way?},} \apjl, 943, L12,
  \dodoi{10.3847/2041-8213/acad82}

% type= article
\bibitem[{K. {Kunnumkai} {et~al.}(2026){Kunnumkai}, {Palmese}, {O'Connor},
  {Farah}, \& {Magana Hernandez}}]{Kunnumkai+2026}
{Kunnumkai}, K., {Palmese}, A., {O'Connor}, B., {Farah}, A., \& {Magana
  Hernandez}, I. 2026, \bibinfo{title}{{Wide Jets or Low Rates: Reconciling
  Short GRB and Gravitational-Wave Neutron Star Merger Rates},} arXiv e-prints,
  arXiv:2604.05046.
\newblock \doarXiv{2604.05046}

% type= article
\bibitem[{J.~M. {Lattimer} \& D.~N. {Schramm}(1974){Lattimer} \&
  {Schramm}}]{Lattimer&Schramm1974}
{Lattimer}, J.~M., \& {Schramm}, D.~N. 1974,
  \bibinfo{title}{{Black-Hole-Neutron-Star Collisions},} \apjl, 192, L145,
  \dodoi{10.1086/181612}

% type= article
\bibitem[{G. {Limberg} {et~al.}(2024){Limberg}, {Ji}, {Naidu}, {Chiti},
  {Rossi}, {Usman}, {Ting}, {Zaritsky}, {Bonaca}, {Borbolato}, {Speagle},
  {Chandra}, \& {Conroy}}]{Limberg+2024}
{Limberg}, G., {Ji}, A.~P., {Naidu}, R.~P., {et~al.} 2024,
  \bibinfo{title}{{Extending the chemical reach of the H3 survey: detailed
  abundances of the dwarf-galaxy stellar stream Wukong/LMS-1<SUP></SUP>},}
  \mnras, 530, 2512, \dodoi{10.1093/mnras/stae969}

% type= article
\bibitem[{A.~G. {Lyne} \& D.~R. {Lorimer}(1994){Lyne} \&
  {Lorimer}}]{Lyne&Lorimer1994}
{Lyne}, A.~G., \& {Lorimer}, D.~R. 1994, \bibinfo{title}{{High birth velocities
  of radio pulsars},} \nat, 369, 127, \dodoi{10.1038/369127a0}

% type= article
\bibitem[{D. {Maoz} \& E. {Nakar}(2024){Maoz} \& {Nakar}}]{Maoz&Nakar2024}
{Maoz}, D., \& {Nakar}, E. 2024, \bibinfo{title}{{The neutron-star merger
  delay-time distribution, r-process ``knees'', and the metal budget of the
  Galaxy},} arXiv e-prints, arXiv:2406.08630, \dodoi{10.48550/arXiv.2406.08630}

% type= article
\bibitem[{T. {Matsuno} {et~al.}(2021){Matsuno}, {Hirai}, {Tarumi},
  {Hotokezaka}, {Tanaka}, \& {Helmi}}]{Matsuno+2021}
{Matsuno}, T., {Hirai}, Y., {Tarumi}, Y., {et~al.} 2021,
  \bibinfo{title}{{R-process enhancements of Gaia-Enceladus in GALAH DR3},}
  \aap, 650, A110, \dodoi{10.1051/0004-6361/202040227}

% type= article
\bibitem[{K.~S. {McCarthy} {et~al.}(2020){McCarthy}, {Zheng}, \&
  {Ramirez-Ruiz}}]{McCarthy+2020}
{McCarthy}, K.~S., {Zheng}, Z., \& {Ramirez-Ruiz}, E. 2020,
  \bibinfo{title}{{Constraining delay time distribution of binary neutron star
  mergers from host galaxy properties},} \mnras, 499, 5220,
  \dodoi{10.1093/mnras/staa3206}

% type= article
\bibitem[{A. {McWilliam} {et~al.}(1995){McWilliam}, {Preston}, {Sneden}, \&
  {Searle}}]{McWilliam+1995}
{McWilliam}, A., {Preston}, G.~W., {Sneden}, C., \& {Searle}, L. 1995,
  \bibinfo{title}{{Spectroscopic Analysis of 33 of the Most Metal Poor Stars.
  II.},} \aj, 109, 2757, \dodoi{10.1086/117486}

% type= article
\bibitem[{B.~D. {Metzger}(2017){Metzger}}]{Metzger2017}
{Metzger}, B.~D. 2017, \bibinfo{title}{{Welcome to the Multi-Messenger Era!
  Lessons from a Neutron Star Merger and the Landscape Ahead},} arXiv e-prints,
  arXiv:1710.05931, \dodoi{10.48550/arXiv.1710.05931}

% type= article
\bibitem[{B.~D. {Metzger} {et~al.}(2010){Metzger}, {Mart{\'\i}nez-Pinedo},
  {Darbha}, {Quataert}, {Arcones}, {Kasen}, {Thomas}, {Nugent}, {Panov}, \&
  {Zinner}}]{Metzger+2010}
{Metzger}, B.~D., {Mart{\'\i}nez-Pinedo}, G., {Darbha}, S., {et~al.} 2010,
  \bibinfo{title}{{Electromagnetic counterparts of compact object mergers
  powered by the radioactive decay of r-process nuclei},} \mnras, 406, 2650,
  \dodoi{10.1111/j.1365-2966.2010.16864.x}

% type= article
\bibitem[{R.~P. {Naidu} {et~al.}(2022){Naidu}, {Ji}, {Conroy}, {Bonaca},
  {Ting}, {Zaritsky}, {van Son}, {Broekgaarden}, {Tacchella}, {Chandra},
  {Caldwell}, {Cargile}, \& {Speagle}}]{Naidu+2022}
{Naidu}, R.~P., {Ji}, A.~P., {Conroy}, C., {et~al.} 2022,
  \bibinfo{title}{{Evidence from Disrupted Halo Dwarfs that r-process
  Enrichment via Neutron Star Mergers is Delayed by {\ensuremath{\gtrsim}}500
  Myr},} \apjl, 926, L36, \dodoi{10.3847/2041-8213/ac5589}

% type= article
\bibitem[{A. {Nair} \& S. {Stevenson}(2025){Nair} \& {Stevenson}}]{Nair+2025}
{Nair}, A., \& {Stevenson}, S. 2025, \bibinfo{title}{{Formation of heavy double
  neutron stars ─ I. Eddington-limited accretion for a 1.4
  M$_{{\ensuremath{\odot}}}$ neutron star at solar metallicity},} \mnras, 543,
  233, \dodoi{10.1093/mnras/staf1397}

% type= article
\bibitem[{E. {Nakar} {et~al.}(2006){Nakar}, {Gal-Yam}, \& {Fox}}]{Nakar+2006}
{Nakar}, E., {Gal-Yam}, A., \& {Fox}, D.~B. 2006, \bibinfo{title}{{The Local
  Rate and the Progenitor Lifetimes of Short-Hard Gamma-Ray Bursts: Synthesis
  and Predictions for the Laser Interferometer Gravitational-Wave
  Observatory},} \apj, 650, 281, \dodoi{10.1086/505855}

% type= article
\bibitem[{A.~E. {Nugent} {et~al.}(2024){Nugent}, {Fong}, {Castrejon}, {Leja},
  {Zevin}, \& {Ji}}]{Nugent+2024}
{Nugent}, A.~E., {Fong}, W.-f., {Castrejon}, C., {et~al.} 2024,
  \bibinfo{title}{{A Population of Short-duration Gamma-Ray Bursts with Dwarf
  Host Galaxies},} \apj, 962, 5, \dodoi{10.3847/1538-4357/ad17c0}

% type= article
\bibitem[{A.~E. {Nugent} {et~al.}(2025){Nugent}, {Ji}, {Fong}, {Shah}, \& {van
  de Voort}}]{Nugent+2025}
{Nugent}, A.~E., {Ji}, A.~P., {Fong}, W.-f., {Shah}, H., \& {van de Voort}, F.
  2025, \bibinfo{title}{{Where Has All the R-process Gone? Timescales for
  Gamma-Ray Burst Kilonovae to Enrich Their Host Galaxies},} \apj, 982, 144,
  \dodoi{10.3847/1538-4357/adbb6a}

% type= article
\bibitem[{A.~E. {Nugent} {et~al.}(2022){Nugent}, {Fong}, {Dong}, {Leja},
  {Berger}, {Zevin}, {Chornock}, {Cobb}, {Kelley}, {Kilpatrick}, {Levan},
  {Margutti}, {Paterson}, {Perley}, {Escorial}, {Smith}, \&
  {Tanvir}}]{Nugent+2022}
{Nugent}, A.~E., {Fong}, W.-F., {Dong}, Y., {et~al.} 2022,
  \bibinfo{title}{{Short GRB Host Galaxies. II. A Legacy Sample of Redshifts,
  Stellar Population Properties, and Implications for Their Neutron Star Merger
  Origins},} \apj, 940, 57, \dodoi{10.3847/1538-4357/ac91d1}

% type= article
\bibitem[{T. {Nugis} \& H.~J.~G.~L.~M. {Lamers}(2000){Nugis} \&
  {Lamers}}]{Nugis&Lamers2000}
{Nugis}, T., \& {Lamers}, H.~J.~G.~L.~M. 2000, \bibinfo{title}{{Mass-loss rates
  of Wolf-Rayet stars as a function of stellar parameters},} \aap, 360, 227

% type= article
\bibitem[{H. {Okada} {et~al.}(2026){Okada}, {Aoki}, {Tominaga}, \&
  {Honda}}]{Okada+2026}
{Okada}, H., {Aoki}, W., {Tominaga}, N., \& {Honda}, S. 2026,
  \bibinfo{title}{{SMSS J022423.27{\ensuremath{-}}573705.1: An Extremely
  Metal-poor Star with the Most Pronounced Weak r-process Signature},} \apj,
  997, 119, \dodoi{10.3847/1538-4357/ae231c}

% type= misc
\bibitem[{J.~P. Ostriker(1973)Ostriker}]{Ostriker1973}
Ostriker, J.~P. 1973, private communication to Paczynski

% type= inproceedings
\bibitem[{B. {Paczynski}(1976){Paczynski}}]{Paczynski+1976}
{Paczynski}, B. 1976, \bibinfo{title}{{Common Envelope Binaries},} in Structure
  and Evolution of Close Binary Systems, ed. P.~{Eggleton}, S.~{Mitton}, \&
  J.~{Whelan}, Vol.~73, 75

% type=
\bibitem[{T. pandas~development team(2020)pandas~development team}]{pandas}
pandas~development team, T. 2020, pandas-dev/pandas: Pandas, latest Zenodo,
  \dodoi{10.5281/zenodo.3509134}

% type= article
\bibitem[{B. {Paxton} {et~al.}(2011){Paxton}, {Bildsten}, {Dotter}, {Herwig},
  {Lesaffre}, \& {Timmes}}]{Paxton+2011}
{Paxton}, B., {Bildsten}, L., {Dotter}, A., {et~al.} 2011,
  \bibinfo{title}{{Modules for Experiments in Stellar Astrophysics (MESA)},}
  \apjs, 192, 3, \dodoi{10.1088/0067-0049/192/1/3}

% type= article
\bibitem[{B. {Paxton} {et~al.}(2013){Paxton}, {Cantiello}, {Arras}, {Bildsten},
  {Brown}, {Dotter}, {Mankovich}, {Montgomery}, {Stello}, {Timmes}, \&
  {Townsend}}]{Paxton+2013}
{Paxton}, B., {Cantiello}, M., {Arras}, P., {et~al.} 2013,
  \bibinfo{title}{{Modules for Experiments in Stellar Astrophysics (MESA):
  Planets, Oscillations, Rotation, and Massive Stars},} \apjs, 208, 4,
  \dodoi{10.1088/0067-0049/208/1/4}

% type= article
\bibitem[{B. {Paxton} {et~al.}(2015){Paxton}, {Marchant}, {Schwab}, {Bauer},
  {Bildsten}, {Cantiello}, {Dessart}, {Farmer}, {Hu}, {Langer}, {Townsend},
  {Townsley}, \& {Timmes}}]{Paxton+2015}
{Paxton}, B., {Marchant}, P., {Schwab}, J., {et~al.} 2015,
  \bibinfo{title}{{Modules for Experiments in Stellar Astrophysics (MESA):
  Binaries, Pulsations, and Explosions},} \apjs, 220, 15,
  \dodoi{10.1088/0067-0049/220/1/15}

% type= article
\bibitem[{B. {Paxton} {et~al.}(2018){Paxton}, {Schwab}, {Bauer}, {Bildsten},
  {Blinnikov}, {Duffell}, {Farmer}, {Goldberg}, {Marchant}, {Sorokina},
  {Thoul}, {Townsend}, \& {Timmes}}]{Paxton+2018}
{Paxton}, B., {Schwab}, J., {Bauer}, E.~B., {et~al.} 2018,
  \bibinfo{title}{{Modules for Experiments in Stellar Astrophysics (MESA):
  Convective Boundaries, Element Diffusion, and Massive Star Explosions},}
  \apjs, 234, 34, \dodoi{10.3847/1538-4365/aaa5a8}

% type= article
\bibitem[{B. {Paxton} {et~al.}(2019){Paxton}, {Smolec}, {Schwab}, {Gautschy},
  {Bildsten}, {Cantiello}, {Dotter}, {Farmer}, {Goldberg}, {Jermyn}, {Kanbur},
  {Marchant}, {Thoul}, {Townsend}, {Wolf}, {Zhang}, \& {Timmes}}]{Paxton+2019}
{Paxton}, B., {Smolec}, R., {Schwab}, J., {et~al.} 2019,
  \bibinfo{title}{{Modules for Experiments in Stellar Astrophysics (MESA):
  Pulsating Variable Stars, Rotation, Convective Boundaries, and Energy
  Conservation},} \apjs, 243, 10, \dodoi{10.3847/1538-4365/ab2241}

% type= article
\bibitem[{F. Pedregosa {et~al.}(2011)Pedregosa, Varoquaux, Gramfort, Michel,
  Thirion, Grisel, Blondel, Prettenhofer, Weiss, Dubourg, Vanderplas, Passos,
  Cournapeau, Brucher, Perrot, \& Duchesnay}]{scikit-learn}
Pedregosa, F., Varoquaux, G., Gramfort, A., {et~al.} 2011,
  \bibinfo{title}{Scikit-learn: Machine Learning in {P}ython,} Journal of
  Machine Learning Research, 12, 2825

% type= article
\bibitem[{P.~C. {Peters}(1964){Peters}}]{Peters1964}
{Peters}, P.~C. 1964, \bibinfo{title}{{Gravitational Radiation and the Motion
  of Two Point Masses},} Physical Review, 136, 1224,
  \dodoi{10.1103/PhysRev.136.B1224}

% type= article
\bibitem[{M. {Pracchia} \& O. {Sharan Salafia}(2026){Pracchia} \& {Sharan
  Salafia}}]{Pracchia&Salafia2026}
{Pracchia}, M., \& {Sharan Salafia}, O. 2026, \bibinfo{title}{{Short gamma-ray
  burst progenitors have short delay times},} arXiv e-prints, arXiv:2601.03861,
  \dodoi{10.48550/arXiv.2601.03861}

% type= article
\bibitem[{V. {Ramachandran} {et~al.}(2024){Ramachandran}, {Sander}, {Pauli},
  {Klencki}, {Backs}, {Tramper}, {Bernini-Peron}, {Crowther}, {Hamann},
  {Ignace}, {Kuiper}, {Oey}, {Oskinova}, {Shenar}, {Todt}, {Vink}, {Wang},
  {Wofford}, \& {the XShootU Collaboration}}]{Ramachandran+2024}
{Ramachandran}, V., {Sander}, A.~A.~C., {Pauli}, D., {et~al.} 2024,
  \bibinfo{title}{{X-Shooting ULLYSES: Massive stars at low metallicity: VIII.
  Stellar and wind parameters of newly revealed stripped stars in Be
  binaries},} \aap, 692, A90, \dodoi{10.1051/0004-6361/202449665}

% type= article
\bibitem[{H. {Reggiani} {et~al.}(2021){Reggiani}, {Schlaufman}, {Casey},
  {Simon}, \& {Ji}}]{Reggiani+2021}
{Reggiani}, H., {Schlaufman}, K.~C., {Casey}, A.~R., {Simon}, J.~D., \& {Ji},
  A.~P. 2021, \bibinfo{title}{{The Most Metal-poor Stars in the Magellanic
  Clouds Are r-process Enhanced},} \aj, 162, 229,
  \dodoi{10.3847/1538-3881/ac1f9a}

% type= article
\bibitem[{I.~U. {Roederer} {et~al.}(2016){Roederer}, {Mateo}, {Bailey},
  {Spencer}, {Crane}, \& {Shectman}}]{Roederer+2016}
{Roederer}, I.~U., {Mateo}, M., {Bailey}, J.~I., {et~al.} 2016,
  \bibinfo{title}{{Detailed chemical abundances in NGC 5824: another metal-poor
  globular cluster with internal heavy element abundance variations},} \mnras,
  455, 2417, \dodoi{10.1093/mnras/stv2462}

% type= article
\bibitem[{I.~U. {Roederer} \& C. {Sneden}(2011){Roederer} \&
  {Sneden}}]{Roederer+2011}
{Roederer}, I.~U., \& {Sneden}, C. 2011, \bibinfo{title}{{Heavy-element
  Dispersion in the Metal-poor Globular Cluster M92},} \aj, 142, 22,
  \dodoi{10.1088/0004-6256/142/1/22}

% type= article
\bibitem[{M. {Safarzadeh} {et~al.}(2019){Safarzadeh}, {Ramirez-Ruiz},
  {Andrews}, {Macias}, {Fragos}, \& {Scannapieco}}]{Safarzadeh+2019}
{Safarzadeh}, M., {Ramirez-Ruiz}, E., {Andrews}, J.~J., {et~al.} 2019,
  \bibinfo{title}{{r-process Enrichment of the Ultra-faint Dwarf Galaxies by
  Fast-merging Double-neutron Stars},} \apj, 872, 105,
  \dodoi{10.3847/1538-4357/aafe0e}

% type= article
\bibitem[{M. {Saleem} {et~al.}(2025){Saleem}, {Chen}, {Siegel}, {Landry},
  {Read}, \& {Wang}}]{Saleem+2025}
{Saleem}, M., {Chen}, H.-Y., {Siegel}, D.~M., {et~al.} 2025,
  \bibinfo{title}{{Mergers Fall Short: Non-merger Channels Required for
  Galactic Heavy Element Production},} arXiv e-prints, arXiv:2508.06020,
  \dodoi{10.48550/arXiv.2508.06020}

% type= article
\bibitem[{D. {Sanyal} {et~al.}(2017){Sanyal}, {Langer}, {Sz{\'e}csi}, {-C
  Yoon}, \& {Grassitelli}}]{Sanyal+2017}
{Sanyal}, D., {Langer}, N., {Sz{\'e}csi}, D., {-C Yoon}, S., \& {Grassitelli},
  L. 2017, \bibinfo{title}{{Metallicity dependence of envelope inflation in
  massive stars},} \aap, 597, A71, \dodoi{10.1051/0004-6361/201629612}

% type= article
\bibitem[{M.~D. {Shetrone} {et~al.}(2001){Shetrone}, {C{\^o}t{\'e}}, \&
  {Sargent}}]{Shetrone+2001}
{Shetrone}, M.~D., {C{\^o}t{\'e}}, P., \& {Sargent}, W.~L.~W. 2001,
  \bibinfo{title}{{Abundance Patterns in the Draco, Sextans, and Ursa Minor
  Dwarf Spheroidal Galaxies},} \apj, 548, 592, \dodoi{10.1086/319022}

% type= article
\bibitem[{D.~M. {Siegel}(2019){Siegel}}]{Siegel2019}
{Siegel}, D.~M. 2019, \bibinfo{title}{{GW170817 -the first observed neutron
  star merger and its kilonova: Implications for the astrophysical site of the
  r-process},} European Physical Journal A, 55, 203,
  \dodoi{10.1140/epja/i2019-12888-9}

% type= article
\bibitem[{P. {Simonetti} {et~al.}(2019){Simonetti}, {Matteucci}, {Greggio}, \&
  {Cescutti}}]{Simonetti+2019}
{Simonetti}, P., {Matteucci}, F., {Greggio}, L., \& {Cescutti}, G. 2019,
  \bibinfo{title}{{A new delay time distribution for merging neutron stars
  tested against Galactic and cosmic data},} \mnras, 486, 2896,
  \dodoi{10.1093/mnras/stz991}

% type= article
\bibitem[{{\'A}. {Sk{\'u}lad{\'o}ttir} \& S.
  {Salvadori}(2020){Sk{\'u}lad{\'o}ttir} \&
  {Salvadori}}]{Skuladottir&Salvadori2020}
{Sk{\'u}lad{\'o}ttir}, {\'A}., \& {Salvadori}, S. 2020,
  \bibinfo{title}{{Evidence for {\ensuremath{\gtrsim}}4 Gyr timescales of
  neutron star mergers from Galactic archaeology},} \aap, 634, L2,
  \dodoi{10.1051/0004-6361/201937293}

% type= article
\bibitem[{C. {Sneden} {et~al.}(1994){Sneden}, {Preston}, {McWilliam}, \&
  {Searle}}]{Sneden+1994}
{Sneden}, C., {Preston}, G.~W., {McWilliam}, A., \& {Searle}, L. 1994,
  \bibinfo{title}{{Ultra--Metal-poor Halo Stars: The Remarkable Spectrum of CS
  22892-052},} \apjl, 431, L27, \dodoi{10.1086/187464}

% type= article
\bibitem[{N. {Soker}(2026){Soker}}]{Soker2026}
{Soker}, N. 2026, \bibinfo{title}{{The early r-process nucleosynthesis
  scenarios},} arXiv e-prints, arXiv:2601.15187,
  \dodoi{10.48550/arXiv.2601.15187}

% type= article
\bibitem[{H.~F. {Stevance} {et~al.}(2023){Stevance}, {Eldridge}, {Stanway},
  {Lyman}, {McLeod}, \& {Levan}}]{Stevance+2023}
{Stevance}, H.~F., {Eldridge}, J.~J., {Stanway}, E.~R., {et~al.} 2023,
  \bibinfo{title}{{End-to-end study of the host galaxy and genealogy of the
  first binary neutron star merger},} Nature Astronomy, 7, 444,
  \dodoi{10.1038/s41550-022-01873-y}

% type= article
\bibitem[{T. {Sukhbold} {et~al.}(2016){Sukhbold}, {Ertl}, {Woosley}, {Brown},
  \& {Janka}}]{Sukhbold+2016}
{Sukhbold}, T., {Ertl}, T., {Woosley}, S.~E., {Brown}, J.~M., \& {Janka}, H.~T.
  2016, \bibinfo{title}{{Core-collapse Supernovae from 9 to 120 Solar Masses
  Based on Neutrino-powered Explosions},} \apj, 821, 38,
  \dodoi{10.3847/0004-637X/821/1/38}

% type= article
\bibitem[{E. {Symbalisty} \& D.~N. {Schramm}(1982){Symbalisty} \&
  {Schramm}}]{Symbalisty&Schramm1982}
{Symbalisty}, E., \& {Schramm}, D.~N. 1982, \bibinfo{title}{{Neutron Star
  Collisions and the r-Process},} \aplett, 22, 143

% type= article
\bibitem[{T.~M. {Tauris} {et~al.}(2015){Tauris}, {Langer}, \&
  {Podsiadlowski}}]{Tauris+2015}
{Tauris}, T.~M., {Langer}, N., \& {Podsiadlowski}, P. 2015,
  \bibinfo{title}{{Ultra-stripped supernovae: progenitors and fate},} \mnras,
  451, 2123, \dodoi{10.1093/mnras/stv990}

% type= article
\bibitem[{T.~M. {Tauris} {et~al.}(2017){Tauris}, {Kramer}, {Freire}, {Wex},
  {Janka}, {Langer}, {Podsiadlowski}, {Bozzo}, {Chaty}, {Kruckow}, {van den
  Heuvel}, {Antoniadis}, {Breton}, \& {Champion}}]{Tauris+2017}
{Tauris}, T.~M., {Kramer}, M., {Freire}, P.~C.~C., {et~al.} 2017,
  \bibinfo{title}{{Formation of Double Neutron Star Systems},} \apj, 846, 170,
  \dodoi{10.3847/1538-4357/aa7e89}

% type= article
\bibitem[{A. {Vigna-G{\'o}mez} {et~al.}(2022){Vigna-G{\'o}mez}, {Wassink},
  {Klencki}, {Istrate}, {Nelemans}, \& {Mandel}}]{Vigna-Gomez+2022}
{Vigna-G{\'o}mez}, A., {Wassink}, M., {Klencki}, J., {et~al.} 2022,
  \bibinfo{title}{{Stellar response after stripping as a model for
  common-envelope outcomes},} \mnras, 511, 2326, \dodoi{10.1093/mnras/stac237}

% type= article
\bibitem[{A. {Vigna-G{\'o}mez} {et~al.}(2018){Vigna-G{\'o}mez}, {Neijssel},
  {Stevenson}, {Barrett}, {Belczynski}, {Justham}, {de Mink}, {M{\"u}ller},
  {Podsiadlowski}, {Renzo}, {Sz{\'e}csi}, \& {Mandel}}]{Vigna-Gomez+2018}
{Vigna-G{\'o}mez}, A., {Neijssel}, C.~J., {Stevenson}, S., {et~al.} 2018,
  \bibinfo{title}{{On the formation history of Galactic double neutron stars},}
  \mnras, 481, 4009, \dodoi{10.1093/mnras/sty2463}

% type= article
\bibitem[{J.~S. {Vink}(2017){Vink}}]{Vink2017}
{Vink}, J.~S. 2017, \bibinfo{title}{{Winds from stripped low-mass helium stars
  and Wolf-Rayet stars},} \aap, 607, L8, \dodoi{10.1051/0004-6361/201731902}

% type= article
\bibitem[{P. Virtanen {et~al.}(2020)Virtanen, Gommers, Oliphant, Haberland,
  Reddy, Cournapeau, Burovski, Peterson, Weckesser, Bright, {van der Walt},
  Brett, Wilson, Millman, Mayorov, Nelson, Jones, Kern, Larson, Carey, Polat,
  Feng, Moore, {VanderPlas}, Laxalde, Perktold, Cimrman, Henriksen, Quintero,
  Harris, Archibald, Ribeiro, Pedregosa, {van Mulbregt}, \& {SciPy 1.0
  Contributors}}]{scipy}
Virtanen, P., Gommers, R., Oliphant, T.~E., {et~al.} 2020,
  \bibinfo{title}{{{SciPy} 1.0: Fundamental Algorithms for Scientific Computing
  in Python},} Nature Methods, 17, 261, \dodoi{10.1038/s41592-019-0686-2}

% type= article
\bibitem[{D. {Wanderman} \& T. {Piran}(2015){Wanderman} \&
  {Piran}}]{Wanderman&Piran2015}
{Wanderman}, D., \& {Piran}, T. 2015, \bibinfo{title}{{The rate, luminosity
  function and time delay of non-Collapsar short GRBs},} \mnras, 448, 3026,
  \dodoi{10.1093/mnras/stv123}

% type= article
\bibitem[{C. {Xin} {et~al.}(2022){Xin}, {Renzo}, \& {Metzger}}]{Xin+2022}
{Xin}, C., {Renzo}, M., \& {Metzger}, B.~D. 2022, \bibinfo{title}{{Dissecting
  the microphysics behind the metallicity-dependence of massive stars radii},}
  \mnras, 516, 5816, \dodoi{10.1093/mnras/stac2551}

% type= article
\bibitem[{M. {Zevin} {et~al.}(2019){Zevin}, {Kremer}, {Siegel}, {Coughlin},
  {Tsang}, {Berry}, \& {Kalogera}}]{Zevin+2019}
{Zevin}, M., {Kremer}, K., {Siegel}, D.~M., {et~al.} 2019, \bibinfo{title}{{Can
  Neutron-star Mergers Explain the r-process Enrichment in Globular
  Clusters?},} \apj, 886, 4, \dodoi{10.3847/1538-4357/ab498b}

% type= article
\bibitem[{M. {Zevin} {et~al.}(2022){Zevin}, {Nugent}, {Adhikari}, {Fong},
  {Holz}, \& {Kelley}}]{Zevin+2022}
{Zevin}, M., {Nugent}, A.~E., {Adhikari}, S., {et~al.} 2022,
  \bibinfo{title}{{Observational Inference on the Delay Time Distribution of
  Short Gamma-Ray Bursts},} \apjl, 940, L18, \dodoi{10.3847/2041-8213/ac91cd}

% type= article
\bibitem[{Z. {Zheng} \& E. {Ramirez-Ruiz}(2007){Zheng} \&
  {Ramirez-Ruiz}}]{Zheng&Ramirez-Ruiz2007}
{Zheng}, Z., \& {Ramirez-Ruiz}, E. 2007, \bibinfo{title}{{Deducing the Lifetime
  of Short Gamma-Ray Burst Progenitors from Host Galaxy Demography},} \apj,
  665, 1220, \dodoi{10.1086/519544}

\end{thebibliography}
\bibliographystyle{aasjournalv7}

\end{document}